\documentclass[11pt,a4paper]{article}

\input{glyphtounicode}
\usepackage[utf8]{inputenc}
\usepackage[T1]{fontenc}
\usepackage{amsmath,amssymb,amsfonts}
\usepackage{graphicx}
\usepackage{booktabs}
\usepackage[margin=1in]{geometry}
\usepackage{braket}
\usepackage{xcolor}
\usepackage{float}
\usepackage{hyperref}

\title{Quantum Amplitude Estimation for Catastrophe Insurance\\
Tail-Risk Pricing:\\
Empirical Convergence and NISQ Noise Analysis}

\author{Alexis Kirke}

\date{}

\begin{document}

\maketitle


\begin{abstract}
Classical Monte Carlo methods for pricing catastrophe insurance tail risk converge at $O(1/\sqrt{N})$, requiring large simulation budgets to resolve upper-tail percentiles of the loss distribution. This sample-sparsity problem can lead to AI models trained on impoverished tail data, producing poorly calibrated risk estimates where insolvency risk is greatest \cite{Embrechts1997}. Quantum Amplitude Estimation (QAE), following Montanaro \cite{Montanaro2015}, achieves convergence approaching $O(1/N)$ in oracle queries---a quadratic speedup that, at scale, would enable high-resolution tail estimation within practical budgets.

We validate this advantage empirically using a Qiskit Aer simulator with genuine Grover amplification. A complete pipeline encodes fitted lognormal catastrophe distributions into quantum oracles via amplitude encoding, producing small readout probabilities ($P(\ket{1}) \sim 0.002$--$0.05$) that enable safe Grover amplification with up to $k = 16$ iterations.

Seven experiments on synthetic and real (NOAA Storm Events, 58,028 records) data yield three main findings. (i)~\textbf{Oracle-model advantage}: When quantum AE and classical MC operate on the same discretised distribution---the fair comparison in the oracle model---QAE achieves 1.5--2.5$\times$ lower RMSE on synthetic data (Experiments~1 and~3) and 2--2.5$\times$ on real data (Experiments~4 and~7), with convergence consistent with the quadratic query advantage. (ii)~\textbf{Strong classical baselines win when analytical access is available}: Conditional tail MC, importance sampling, and quasi-Monte Carlo with Sobol sequences (which exploit the closed-form lognormal CDF/inverse CDF) dramatically outperform QAE (Experiment~5); QMC's near-$O(1/N)$ convergence matches QAE's theoretical rate in one dimension. On an empirical PMF with no parametric form (Experiment~7), these methods offer no direct advantage without extra modelling, and the oracle-model comparison applies. (iii)~\textbf{Discretisation, not estimation, is the current bottleneck}: A qubit sweep ($n = 3$--$8$, 8--256 bins) shows that tail truncation under coarse equal-width binning dominates overall error (Experiment~6); NISQ noise destroys the advantage entirely (Experiment~2). All error decompositions are reported against both analytic and exact-on-bins ground truths.
\end{abstract}


\section{Introduction}
\label{sec:introduction}

Catastrophe insurance---covering extreme events such as hurricanes, earthquakes, and floods---occupies a unique position in actuarial science. Unlike high-frequency, low-severity lines such as motor or household contents, catastrophe risk is dominated by the behaviour of the loss distribution's upper tail. A reinsurer writing an excess-of-loss contract at the 95th or 97th percentile of the aggregate loss distribution needs to estimate the expected payoff $E[\max(0, X - M)]$ at the attachment point $M$, and small errors in this estimate can translate directly into mispriced risk and, ultimately, insolvency \cite{Embrechts1997, Grossi2005}.

The industry-standard approach to this estimation problem is Monte Carlo (MC) simulation. A catastrophe model---typically a multi-stage pipeline combining hazard generation (e.g.\ wind-field simulation), vulnerability modelling (structural damage as a function of hazard intensity), and financial loss aggregation---is executed many thousands of times to build up an empirical distribution of losses. The expected excess loss is then computed as a sample average over the simulated tail. The fundamental limitation of this approach is well known: classical MC converges at $O(1/\sqrt{N})$, where $N$ is the number of simulation runs. Achieving one additional decimal digit of precision therefore requires a hundredfold increase in computational budget---a prohibitive cost when each simulation run involves complex physical and financial modelling.

This convergence bottleneck is not merely a computational inconvenience; it has direct consequences for the quality of risk estimates. At the 97th percentile, fewer than 3\% of simulated scenarios contribute to the tail expectation. With a budget of $N = 10{,}000$ runs, only $\sim$300 samples inform the estimate, leading to high variance that propagates through pricing, reserving, and capital allocation decisions. When machine-learning models are trained on such impoverished tail data---an increasingly common practice in modern insurtech---the result is poorly calibrated risk predictions precisely where calibration matters most.

Quantum computing offers a theoretically grounded path to overcoming this convergence barrier. The key algorithmic primitive is \emph{quantum amplitude estimation} (QAE), introduced by Brassard et al.\ \cite{Brassard2002} and analysed in the Monte Carlo context by Montanaro \cite{Montanaro2015}. Where classical MC requires $O(1/\epsilon^2)$ samples to achieve estimation error $\epsilon$, QAE achieves $O(1/\epsilon)$---a quadratic speedup. In practical terms, this means that an estimate requiring one million classical simulation runs could, in principle, be obtained with only one thousand quantum oracle calls.

The application of QAE to financial risk estimation has been explored in a series of influential papers from the quantum finance community, most notably by Woerner and Egger \cite{Woerner2019} on quantum risk analysis, Rebentrost et al.\ \cite{Rebentrost2018} on derivative pricing, Stamatopoulos et al.\ \cite{Stamatopoulos2020} on option pricing, and the landmark resource estimation by Chakrabarti et al.\ \cite{Chakrabarti2021} from Goldman Sachs and IBM. However, this body of work has focused almost exclusively on derivatives pricing in capital markets---options, autocallables, and portfolio risk metrics such as Value-at-Risk (VaR) and Conditional Value-at-Risk (CVaR). The insurance domain, despite sharing the same underlying mathematical structure (estimation of tail expectations under complex loss distributions), has received comparatively little attention.

This paper bridges the gap by applying quantum amplitude estimation to catastrophe insurance tail-risk pricing and providing a rigorous empirical evaluation against both naive and variance-reduced classical baselines. Our contributions are threefold:

\begin{enumerate}
    \item We construct a complete pipeline that encodes fitted lognormal catastrophe distributions into quantum oracles via amplitude encoding, producing readout probabilities in the regime ($P(\ket{1}) \sim 0.002$--$0.05$) where Grover amplification is both safe and effective.
    \item We conduct seven experiments on synthetic and real catastrophe data (58,028 NOAA Storm Events records), providing full error decompositions that separate estimation error from discretisation error---a distinction often overlooked in quantum finance literature.
    \item We compare QAE not only against naive MC but also against conditional tail MC, importance sampling, and quasi-Monte Carlo with Sobol sequences, establishing precisely when and why the quantum advantage applies: in the oracle model (black-box access to the loss distribution), but not when analytical access to the distribution's closed-form CDF is available. We also provide fitted log--log convergence exponents with bootstrap confidence intervals and demonstrate that log-spaced binning can reduce the discretisation bottleneck by $10\times$.
\end{enumerate}

The remainder of this paper is organised as follows. Section~\ref{sec:related_work} reviews the relationship between classical and quantum Monte Carlo methods, surveys the quantum finance literature, and positions our contribution within the catastrophe modelling landscape. Section~\ref{sec:methods} presents the full methodology. Section~\ref{sec:results} reports the results of all seven experiments with detailed error analysis. Section~\ref{sec:conclusions} discusses implications and outlines future work.


\section{Related Work}
\label{sec:related_work}

\subsection{Classical Monte Carlo in Insurance and Finance}

Monte Carlo simulation has been the workhorse of quantitative risk analysis since its formalisation in the mid-twentieth century. In insurance, the method is indispensable for pricing catastrophe risk, where the loss distribution is the output of a complex, multi-stage simulation pipeline---hazard generation, vulnerability modelling, and financial aggregation---that admits no closed-form solution \cite{Grossi2005}. The same is true in derivatives pricing: for path-dependent and multi-asset contracts, MC simulation is often the only viable approach \cite{Glasserman2003}.

The fundamental limitation is convergence. Classical MC estimation of an expectation $E[f(X)]$ from $N$ i.i.d.\ samples converges at rate $O(1/\sqrt{N})$ in root-mean-squared error (RMSE). This rate is a direct consequence of the Central Limit Theorem and cannot be improved for general black-box distributions. For tail-risk estimation, the problem is compounded: if $f(X) = \max(0, X - M)$ and $M$ is at the 97th percentile, then $f(X) = 0$ for 97\% of samples, and the effective sample size for the tail expectation is only $\sim 0.03N$.

Classical variance-reduction techniques---importance sampling, conditional tail sampling, stratified sampling, and control variates---can dramatically improve convergence for specific problem structures. In particular, when the loss distribution has a known parametric form (e.g.\ lognormal, Pareto, or generalised Pareto), conditional tail MC and importance sampling can reduce RMSE by orders of magnitude relative to naive MC. Quasi-Monte Carlo (QMC) methods, which replace pseudo-random samples with low-discrepancy sequences (Sobol, Halton), achieve $O((\log N)^d/N)$ convergence for smooth $d$-dimensional integrands---near-$O(1/N)$ in low dimensions---and are widely used in financial Monte Carlo \cite{Glasserman2003}. QMC's convergence rate theoretically matches that of QAE, making it an essential classical comparator (see Experiment~5). However, all these techniques require analytical access to the distribution's CDF, PDF, or inverse CDF---information that is not available when the distribution is defined implicitly as the output of a catastrophe simulation pipeline.

\subsection{From Classical to Quantum Monte Carlo: The Amplitude Estimation Bridge}

The connection between classical MC and quantum computing rests on a precise mathematical analogy. In classical MC, one estimates $E[f(X)]$ by drawing samples $X_1, \ldots, X_N$ from the distribution of $X$ and computing the sample mean $\hat{\mu} = (1/N) \sum_{i=1}^{N} f(X_i)$. The standard error of $\hat{\mu}$ is $\sigma_f / \sqrt{N}$, where $\sigma_f$ is the standard deviation of $f(X)$.

In quantum amplitude estimation, one encodes the distribution and payoff function into a quantum circuit (the ``oracle'') such that the probability of measuring a designated ancilla qubit in the state $\ket{1}$ equals the quantity to be estimated (suitably normalised). The key insight, due to Brassard et al.\ \cite{Brassard2002}, is that this probability can be amplified and estimated using Grover's search algorithm \cite{Grover1996} as a subroutine. Where each classical MC sample provides $O(1)$ bits of information about the mean, each application of the Grover operator provides $O(1)$ bits of information about the \emph{phase} of the amplitude, enabling estimation at rate $O(1/N)$ in the number of oracle queries $N$---a quadratic improvement.

Montanaro \cite{Montanaro2015} formalised this connection, proving that quantum amplitude estimation achieves a quadratic speedup over classical MC for estimating expectations of bounded functions. The result applies to any distribution that can be prepared as a quantum state and any payoff function that can be implemented as a unitary rotation on an ancilla qubit. Crucially, the speedup holds in the \emph{oracle model}: the quantum algorithm accesses the distribution only through the oracle circuit, just as classical MC accesses it only through samples. When additional structure is available (e.g.\ a closed-form CDF), classical methods such as importance sampling can exploit that structure in ways the oracle model does not capture---a distinction that is central to the interpretation of our results.

More recent work has extended the amplitude estimation framework. Suzuki et al.\ \cite{Suzuki2020} developed amplitude estimation without phase estimation, removing the need for quantum phase estimation circuits and reducing the required circuit depth. This line of work makes QAE more practical for near-term quantum devices, although the fundamental quadratic advantage remains tied to the number of Grover iterations that can be coherently executed.

\subsection{Quantum Monte Carlo in Finance: From Theory to Resource Estimation}

The application of quantum amplitude estimation to financial problems was pioneered by Rebentrost et al.\ \cite{Rebentrost2018}, who showed how probability distributions can be prepared in quantum superposition, payoff functions implemented as quantum circuits, and derivative prices extracted via quantum measurement. Their work established the basic pipeline---state preparation, payoff encoding, amplitude estimation---that all subsequent quantum finance papers have followed.

Woerner and Egger \cite{Woerner2019} applied this framework to risk analysis, demonstrating quantum circuits for estimating VaR and CVaR on small-scale instances. Stamatopoulos et al.\ \cite{Stamatopoulos2020} extended the approach to option pricing, implementing European and Asian option circuits on IBM quantum hardware. These early implementations were proof-of-concept demonstrations on small qubit counts (3--5 qubits), consistent with the scale of experiments in the present paper.

A pivotal contribution came from Chakrabarti et al.\ \cite{Chakrabarti2021}, a collaboration between Goldman Sachs and IBM that provided the first detailed end-to-end resource estimate for quantum advantage in derivative pricing. Working with benchmark derivative contracts (autocallables and Target Accrual Redemption Notes), they estimated that quantum advantage would require approximately 7,500 logical qubits and a T-depth of 46 million, with T-gates running at 10~MHz or faster---specifications far beyond current hardware, but providing a concrete target for the quantum computing roadmap. Their ``re-parameterisation method'' for loading stochastic processes into quantum circuits also addressed a key practical bottleneck in state preparation.

Building on this foundation, Stamatopoulos and Zeng \cite{Stamatopoulos2024} from Goldman Sachs introduced quantum signal processing (QSP) for encoding derivative payoffs, reducing the total T-gate count by approximately $16\times$ and the logical qubit count by approximately $4\times$ relative to previous approaches. This work represents the current state of the art in resource-efficient quantum derivative pricing circuits.

The present paper extends this line of research from derivatives to insurance in two significant ways. First, the loss distributions in catastrophe insurance are fundamentally different from those in derivatives pricing: they are heavy-tailed, right-skewed, and often derived from physical simulation rather than stochastic-process models. Second, we provide a more nuanced comparison with classical baselines than is typical in the quantum finance literature, distinguishing between the oracle-model advantage (which is real) and the end-to-end advantage (which depends on discretisation quality and the availability of analytical structure).

\subsection{Catastrophe Modelling and Tail-Risk Estimation}

Catastrophe modelling emerged as a discipline in the 1990s following a series of unprecedented insurance losses (Hurricane Andrew in 1992, the Northridge earthquake in 1994) that exposed the inadequacy of purely actuarial approaches to pricing extreme events \cite{Grossi2005}. Modern catastrophe models are complex simulation pipelines that incorporate meteorological or seismological hazard models, engineering-based vulnerability functions, and financial loss aggregation with policy terms and reinsurance structures. The computational cost of running these models---typically minutes to hours per scenario, depending on resolution---makes the convergence rate of the sampling method a binding constraint on the quality of risk estimates.

The insurance industry has explored various approaches to improving tail estimation. Extreme value theory (EVT) provides parametric models for the tail (generalised Pareto distribution, peaks-over-threshold methods) that can be fitted to the upper tail of simulated losses \cite{Embrechts1997}. These methods improve estimation efficiency but introduce model risk: the fitted parametric form may not accurately represent the true tail.

Quantum computing offers an orthogonal approach: rather than approximating the model or fitting a parametric tail, QAE can in principle estimate any tail functional directly from the oracle with provably faster convergence. The practical realisation of this advantage depends on several factors that this paper investigates empirically: the quality of the quantum discretisation (how many bins/qubits are needed to capture the tail), the impact of NISQ noise on the amplified circuits, and the relative performance against variance-reduced classical methods that exploit distributional structure.


\section{Methods}
\label{sec:methods}

\subsection{Loss Distribution and Data Pipeline}

The pipeline begins by acquiring empirical loss data to parameterise the catastrophe distribution. Two data sources are supported. For Experiments~1--3, synthetic data is drawn from a heavy-tailed Pareto distribution with shape parameter $\alpha = 1.5$ and scale parameter \$50,000, values chosen to approximate the empirical severity profile of US hurricane property damage \cite{Grossi2005}; 20,000 records are generated. For Experiment~4, the pipeline discovers and downloads NOAA Storm Events detail files \cite{NOAA2024} spanning 2020--2024, extracting all records with property damage $\geq$~\$1,000. The downloader caches files locally and writes a pinned manifest (\texttt{data/cache/noaa\_manifest.json}) recording the exact filenames, base URL, and retrieval date. \emph{All results in this paper are derived from the specific file versions listed in Table~\ref{tab:noaa_files}; the results would change if NOAA updates these files.} The pipeline's dynamic discovery feature is a convenience for users running the code, but the analysis is pinned to the cached snapshot. This yields 58,028 real loss records---a substantially more heavy-tailed distribution ($\hat{\sigma} = 2.02$, $\hat{\mu} = 9.04$, median $\approx$~\$8,465) than the synthetic Pareto ($\hat{\sigma} = 0.67$, $\hat{\mu} = 11.49$, median $\approx$~\$97,203).

Each dataset is fitted to a lognormal distribution via maximum likelihood estimation (with location fixed at zero). We intentionally fit lognormal even to Pareto-generated data to mirror common severity-modelling practice in actuarial science; the quantum/classical comparisons share the same fitted model, so this choice does not bias the estimator comparison. This lognormal fit serves as the continuous distribution from which both the classical Monte Carlo sampler and the quantum oracle are derived.

\begin{table}[t]
\centering
\caption{NOAA Storm Events files used in Experiments~4 and~7. Retrieved 2026-02-27 from ncei.noaa.gov.}
\label{tab:noaa_files}
\begin{tabular}{ll}
\toprule
Filename & Year \\
\midrule
StormEvents\_details-ftp\_v1.0\_d2020\_c20260116.csv.gz & 2020 \\
StormEvents\_details-ftp\_v1.0\_d2021\_c20250520.csv.gz & 2021 \\
StormEvents\_details-ftp\_v1.0\_d2022\_c20250721.csv.gz & 2022 \\
StormEvents\_details-ftp\_v1.0\_d2023\_c20260116.csv.gz & 2023 \\
StormEvents\_details-ftp\_v1.0\_d2024\_c20260116.csv.gz & 2024 \\
\bottomrule
\end{tabular}
\end{table}

\subsection{Quantum State Preparation and Oracle Construction}

The fitted lognormal distribution is discretised into $2^n = 8$ bins ($n = 3$ qubits) spanning the 0.1st to 99.9th percentiles. Each bin $i$ has midpoint $x_i$ and probability mass $p_i = \Pr(X \in \text{bin}_i)$, with probabilities summing to unity by construction. For a given catastrophe threshold $M$ (the reinsurance attachment point), the excess-of-loss payoff is $f(x_i) = \max(0, x_i - M)$, which is normalised to $\tilde{f}(x_i) = f(x_i)/f_{\max} \in [0,1]$, where $f_{\max} = \max_j f(x_j)$.

The quantum oracle $A$ is constructed on $n+1 = 4$ qubits (3 index qubits and 1 ancilla) using amplitude encoding in two stages:

\paragraph{Stage~1: Custom state preparation.} The distribution is loaded directly into qubit amplitudes:
\begin{equation}
\ket{0\ldots 0} \longrightarrow \ket{\psi} = \sum_{i=0}^{N-1} \sqrt{p_i}\,\ket{i},
\label{eq:state_prep}
\end{equation}
using a binary tree of controlled-$R_y$ rotations. At the top level, a single $R_y$ rotation on the most significant qubit splits the total probability mass between the first and second halves of the distribution. At each subsequent level, controlled-$R_y$ gates (controlled on the higher qubits already prepared) further subdivide the mass within each subspace. For $n$ qubits, this requires $2^n - 1$ rotations using only $\{R_y, CR_y, MCR_y, CX, X\}$ gates. Crucially, this avoids Qiskit's \texttt{StatePreparation} instruction, which decomposes into \texttt{cu} (controlled-unitary) gates that crash the Aer simulator when composed into the Grover operator's inverse. The custom state preparation achieves machine-precision fidelity ($< 10^{-15}$ $L_\infty$ error in $|a_i|^2$ vs.\ $p_i$).

\paragraph{Stage~2: Payoff rotation on the ancilla.} For each basis state $\ket{i}$, a multi-controlled $R_y$ rotation is applied to the ancilla qubit with angle $\theta_i = 2\arcsin(\sqrt{\tilde{f}(x_i)})$, so that
\begin{equation}
\ket{i}\ket{0} \longrightarrow \ket{i}\Big(\cos(\theta_i/2)\ket{0} + \sin(\theta_i/2)\ket{1}\Big),
\label{eq:payoff_rotation}
\end{equation}
where $\sin^2(\theta_i/2) = \tilde{f}(x_i)$. The probability of measuring the ancilla in $\ket{1}$ is then
\begin{equation}
P(\ket{1}) = \sum_{i=0}^{N-1} p_i \cdot \tilde{f}(x_i) = \frac{E[\max(0, X - M)]}{f_{\max}},
\label{eq:readout_prob}
\end{equation}
so that the true expected excess loss is recovered by
\begin{equation}
E[\max(0, X - M)] = P(\ket{1}) \times f_{\max}.
\label{eq:recovery}
\end{equation}

The key advantage of amplitude encoding over the uniform-superposition approach (where Hadamard gates prepare $(1/\sqrt{N})\sum\ket{i}$ and both probabilities and payoffs are packed into the rotation angle) is that for tail events, most of the probability mass $p_i$ is concentrated in low-loss bins where $\tilde{f}(x_i) = 0$. This makes $P(\ket{1})$ genuinely small---typically 0.002--0.02 for thresholds at the 90th--97th percentiles---which is precisely the regime where Grover amplification is safe and effective.

\subsection{Grover-Boosted Amplitude Estimation}

After the initial oracle $A$, $k$ iterations of the Grover operator $Q = A\, S_0\, A^\dagger\, S_\chi$ are applied, where $S_\chi$ flips the phase of states with ancilla $= \ket{1}$ (implemented as a $Z$ gate) and $S_0$ flips the phase of the all-zero state (implemented as $X^{\otimes n}\, H\, MCX\, H\, X^{\otimes n}$). The ancilla is then measured over $S$ shots, yielding measured probability $\hat{p}_{\text{meas}}$.

We use the following notation consistently: $P(\ket{1})$ denotes the \emph{true} (circuit-level) probability of measuring the ancilla in $\ket{1}$, while $\hat{p}_{\text{meas}}$ denotes the \emph{measured} (empirical) frequency from a finite number of shots. With $k$ Grover iterations, the noiseless measurement probability is amplified to $\sin^2((2k+1)\theta)$ where $\theta = \arcsin(\sqrt{P(\ket{1})})$, and the measured frequency $\hat{p}_{\text{meas}}$ approximates this. The de-amplification correction
\begin{equation}
\hat{\theta} = \frac{\arcsin(\sqrt{\hat{p}_{\text{meas}}})}{2k+1}, \quad \hat{P} = \sin^2(\hat{\theta})
\label{eq:deamp}
\end{equation}
recovers an estimate of the original $P(\ket{1})$. This is only valid when $(2k+1)\theta < \pi/2$; accordingly, the maximum safe iteration count is
\begin{equation}
k_{\max} = \left\lfloor \frac{\pi/(2\theta) - 1}{2} \right\rfloor,
\label{eq:kmax}
\end{equation}
and all experiments enforce $k \leq k_{\max}$ to prevent angle aliasing.

The Grover operator is built using \texttt{A.to\_gate()} and elementary gates ($H$, $X$, $Z$, $MCX$) that the Aer simulator handles natively after transpilation. For the 95th-percentile threshold with $n=3$ at $k=6$, the full circuit is composed of $(2k+1) = 13$ oracle calls; Table~\ref{tab:circuit_resources} reports the single-oracle depth and two-qubit gate count after decomposition to $\{CX, R_z, SX, X\}$.

\subsection{Noise Model}

For the NISQ noise study, the noiseless Aer backend is replaced with an \texttt{AerSimulator} configured with a depolarising noise model. Three severity presets are tested:

\begin{center}
\begin{tabular}{lccc}
\toprule
Preset & $p_{1q}$ & $p_{2q}$ & $p_{\text{readout}}$ \\
\midrule
Low & 0.001 & 0.01 & 0.005 \\
Medium & 0.005 & 0.05 & 0.02 \\
High & 0.01 & 0.10 & 0.05 \\
\bottomrule
\end{tabular}
\end{center}

Depolarising errors are applied to all single-qubit gates and all two-qubit gates, with symmetric readout error on all qubits. The noise model uses i.i.d.\ depolarising channels and does not capture device-specific effects such as crosstalk, $T_1/T_2$ relaxation, or leakage.

\subsection{Experimental Protocol}

Seven experiments are conducted, all with global random seed 42 for reproducibility. Experiments~1--3 are described below; classical baselines for budget-matched comparison are defined in \S\ref{sec:baselines}; Experiments~4--7 protocols are detailed alongside their results.

\paragraph{Experiment~1: Noiseless Convergence Scaling.} The expected excess loss $E[\max(0, X-M)]$ at the 95th-percentile threshold ($M = \$362{,}700$) is estimated by both classical MC and Grover-amplified quantum methods. Grover iterations are swept from $k=0$ to $k=6$ (within the safe range $k_{\max} = 9$) at a fixed 1,000 shots per quantum run. For each $k$, the total oracle-query budget is $S \times (2k+1)$, and classical MC is given the same total budget as its sample count. Each configuration is repeated 30 times. To avoid conflating discretisation bias with estimation error, we report RMSE against two ground truths: the analytic lognormal excess (\$2,834) and the exact-on-bins value (\$2,330), with discretisation error \$505. Quantum AE and classical MC on the same bins are compared against the exact-on-bins truth (oracle-model comparison); classical MC sampling from the continuous lognormal is compared against the analytic truth.

\paragraph{Experiment~2: NISQ Noise Degradation.} The oracle circuit is executed at $k=3$ Grover iterations (chosen as a moderate amplification depth) at all four noise levels with 8,192 shots, repeated 20 times. This tests noise degradation on a circuit of meaningful depth ($\sim$700 gates).

\paragraph{Experiment~3: Tail-Specific Excess Loss.} The catastrophe threshold $M$ is set to the 90th, 95th, and 97th percentiles of the empirical loss distribution. The quantum estimator uses 8,192 total shots with both $k=0$ (no amplification) and $k_{\text{use}} = \min(k_{\max}, 6)$ Grover iterations (shot count reduced to $S_G = \lfloor 8192/(2k+1)\rfloor$ for the Grover-amplified runs). Two classical baselines receive the same total query budget: (i)~classical MC sampling from the continuous lognormal (RMSE vs analytic truth), and (ii)~classical MC sampling from the same 8-bin discretisation as the quantum circuit (RMSE vs exact-on-bins). All estimators are repeated 30 times.

Experiments~4--7 apply the above methods and additional classical baselines (defined in \S\ref{sec:baselines}) to real NOAA data and empirical PMFs; their protocols are detailed alongside their results.

\subsection{Classical Baselines for Budget-Matched Comparison}
\label{sec:baselines}

Experiments~5 and~7 compare quantum AE against strong classical baselines at strictly equal query/sample budgets. For $X \sim \text{Lognormal}(\mu, \sigma^2)$ with threshold $M$, the following estimators are defined:

\begin{enumerate}
    \item \textbf{Analytic ground truth} (non-sampling reference): $E[(X-M)^+] = e^{\mu+\sigma^2/2}\Phi(d_1) - M[1 - \Phi(d_2)]$ where $d_1 = (\mu + \sigma^2 - \log M)/\sigma$, $d_2 = (\log M - \mu)/\sigma$.
    \item \textbf{Naive MC}: draw $B$ samples from the fitted lognormal; compute $(1/B)\sum \max(0, X_i - M)$.
    \item \textbf{Conditional Tail (CT) MC}: compute $P(X > M)$ analytically, sample $B$ draws from $X \mid X > M$ via truncated inverse CDF, multiply mean excess by $P(X > M)$.
    \item \textbf{Importance Sampling (IS) MC}: apply an exponential tilt $\delta = \max(0, \log M - \mu - \sigma^2/2)$ so the proposal mean $\approx M$; draw $B$ samples from the tilted distribution and reweight.
    \item \textbf{Classical MC on bins}: sample from the same $2^n$-bin discretised distribution as the quantum circuit.
    \item \textbf{Quasi-Monte Carlo (QMC)}: replace pseudo-random samples with a scrambled Sobol low-discrepancy sequence \cite{Glasserman2003}, applying the inverse CDF transform $X = F^{-1}(u)$ to the Sobol points. QMC achieves near-$O(1/N)$ convergence in one dimension and is widely used in financial Monte Carlo.
\end{enumerate}

CT, IS, and QMC all exploit the closed-form lognormal CDF/PDF (or inverse CDF); they cannot be directly applied to non-parametric oracles without an additional density-estimation step (see Experiment~7 discussion). RMSE is computed over 50 repetitions against the analytic ground truth (item~1), which includes discretisation error for the bin-based methods.


\section{Results}
\label{sec:results}

\subsection{Experiment~1: Noiseless Convergence Scaling}

Table~\ref{tab:exp1} presents the core convergence result with full error decomposition. Two ground truths are shown: the analytic lognormal excess (\$2,834) and the exact-on-bins value (\$2,330), with discretisation error \$505. Quantum AE and classical MC on the same bins are measured against the exact-on-bins truth; classical MC sampling from the continuous lognormal is measured against the analytic truth.

\begin{table}[t]
\centering
\caption{Experiment~1: error decomposition at the 95th-percentile threshold ($P(\ket{1}) = 0.0066$, $k_{\max} = 9$, $n_{\text{reps}} = 30$, 1,000 shots/run). Analytic $E[\text{excess}] = \$2{,}834$; exact-on-bins $= \$2{,}330$; disc.\ error $= \$505$. Speedup $= C_{\text{bins}}/\text{Q}$; values ${>}1$ indicate quantum is better.}
\label{tab:exp1}
\begin{tabular}{cccccc}
\toprule
$k$ & Queries & Q (vs bins) & $C_{\text{bins}}$ (vs bins) & Speedup & $C_{\text{cont}}$ (vs anal.) \\
\midrule
0 & 1,000 & 930 & 782 & 0.8$\times$ & 948 \\
1 & 3,000 & 283 & 382 & 1.4$\times$ & 482 \\
2 & 5,000 & 180 & 303 & 1.7$\times$ & 287 \\
3 & 7,000 & 124 & 248 & 2.0$\times$ & 352 \\
4 & 9,000 & 106 & 139 & 1.3$\times$ & 235 \\
5 & 11,000 & 87 & 177 & 2.0$\times$ & 263 \\
6 & 13,000 & 69 & 154 & 2.2$\times$ & 233 \\
\bottomrule
\end{tabular}
\end{table}

In the oracle-model comparison (both methods estimating the same discretised quantity), quantum AE at $k=0$ slightly trails classical MC on bins (\$930 vs \$782, speedup $0.8\times$, i.e.\ quantum is worse), as expected: both are subject to $O(1/\sqrt{N})$ shot noise. As $k$ increases, the quantum RMSE drops from \$930 to \$69, while classical-on-bins drops only from \$782 to \$154, yielding an honest oracle-model speedup of $2.2\times$ at $k=6$. The final column shows that classical MC sampling from the continuous lognormal achieves \$233 RMSE against the analytic truth at $N = 13{,}000$---comparable to classical-on-bins, confirming that discretisation error (\$505) does not dominate the classical-on-bins result at this budget.

\paragraph{Convergence exponents.} To move beyond qualitative claims of ``near-quadratic'' convergence, we fit log--log slopes (RMSE vs.\ total oracle queries) with 95\% bootstrap confidence intervals ($n = 2{,}000$ resamples). At $n=3$ qubits (7 data points, $k=0\ldots 6$), the quantum AE slope is $-1.075$ (95\% CI: $[-1.31, -0.93]$, $R^2 = 0.987$) and classical-on-bins is $-0.539$ (95\% CI: $[-0.67, -0.44]$, $R^2 = 0.978$). To widen the query-budget range, we also ran identical convergence sweeps at $n=8$ qubits (256 bins; see Experiment~6 for circuit metrics). Combining the $n=3$ and $n=8$ data (11 points total) tightens the CIs substantially: quantum slope $-1.074$ (95\% CI: $[-1.17, -1.01]$, $R^2 = 0.988$), classical slope $-0.545$ (95\% CI: $[-0.62, -0.51]$, $R^2 = 0.981$). The combined quantum CI now \emph{excludes} the classical rate entirely, providing stronger evidence for near-$O(1/N)$ convergence than a single qubit count can offer.

\begin{figure}[t]
\centering
\includegraphics[width=\textwidth]{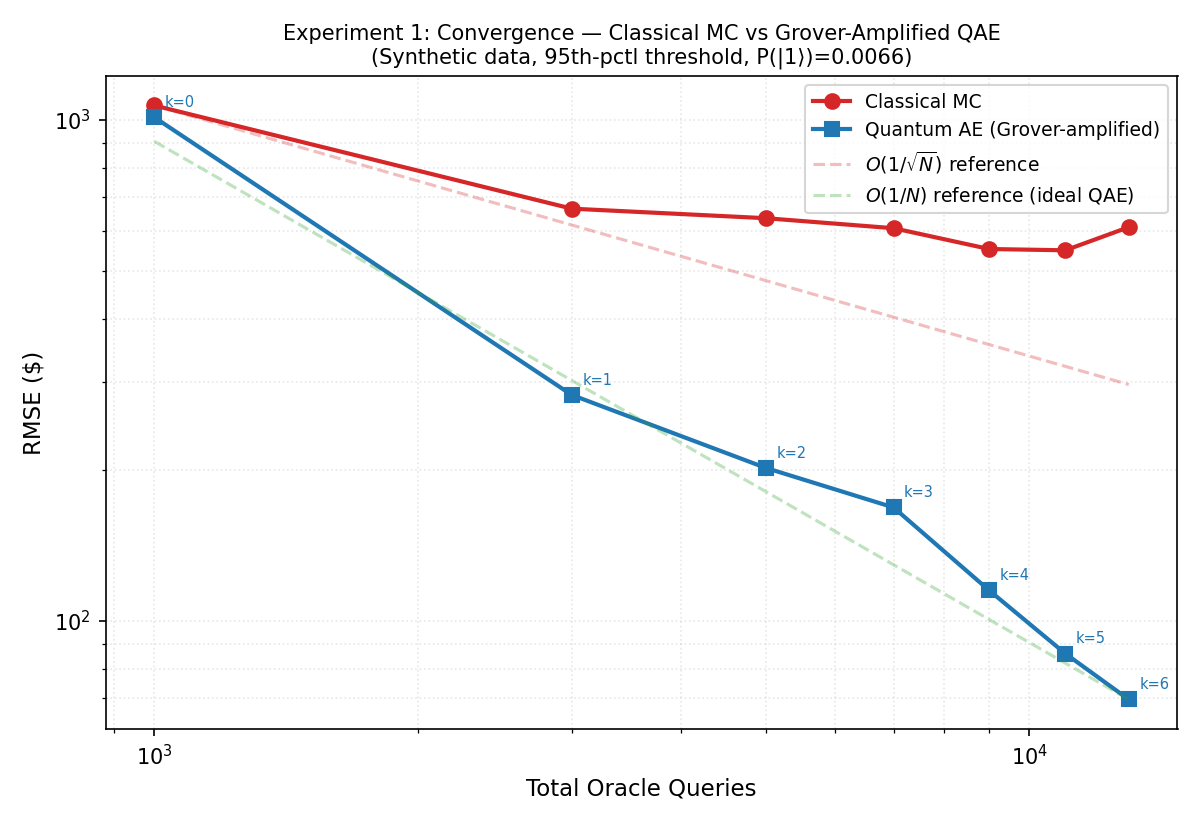}
\caption{Experiment~1: log--log convergence of RMSE vs total oracle queries for quantum AE (Grover-amplified, $k=0\ldots6$) and classical MC on the same 8-bin discretisation. Dashed lines show the $O(1/\sqrt{N})$ and $O(1/N)$ reference slopes.}
\label{fig:exp1}
\end{figure}

Figure~\ref{fig:exp1} shows the log--log convergence. The left panel compares quantum AE and classical MC on the same bins (both vs exact-on-bins); the right panel adds end-to-end context showing the discretisation error floor.

\subsection{Experiment~2: NISQ Noise Degradation}

Table~\ref{tab:exp2} reports QAE accuracy under noise at $k=3$ Grover iterations ($\sim$700-gate circuit), estimating the excess loss at the 95th-percentile threshold (exact value \$2,330).

\begin{table}[t]
\centering
\caption{Experiment~2: QAE accuracy under NISQ noise ($k=3$, 8,192 shots, $n_{\text{reps}}=20$). Exact $E[\text{excess}] = \$2{,}330$.}
\label{tab:exp2}
\begin{tabular}{lcccccc}
\toprule
Noise Level & $p_{1q}$ / $p_{2q}$ & RMSE (\$) & Mean Est.\ (\$) & Std.\ (\$) & Bias (\$) \\
\midrule
Noiseless & 0 / 0 & 32 & 2,332 & 32 & +2 \\
Low & 0.001 / 0.01 & 2,094 & 4,423 & 50 & +2,093 \\
Medium & 0.005 / 0.05 & 2,115 & 4,444 & 62 & +2,114 \\
High & 0.01 / 0.10 & 2,115 & 4,444 & 67 & +2,114 \\
\bottomrule
\end{tabular}
\end{table}

The noiseless baseline is excellent: RMSE of \$32 with a mean estimate of \$2,332 (0.1\% from exact). Under noise, a substantial systematic bias emerges. Even at the ``low'' noise preset ($p_{1q} = 0.001$, $p_{2q} = 0.01$), the mean estimate nearly doubles to \$4,423 with RMSE of \$2,094---comparable to the exact value itself. This dramatic degradation is a direct consequence of the circuit depth required for $k=3$ Grover iterations: the transpiled circuit contains on the order of 50--100 two-qubit gates (Table~\ref{tab:circuit_resources}), giving an expected number of two-qubit gate errors $\sim 50$--$100 \times 0.01 \approx 0.5$--$1$ at the lowest $p_{2q}$, plus additional single-qubit errors across $\sim$700 total gates. Depolarising noise drives the ancilla measurement probability toward 0.5 (the maximally mixed state), and the de-amplification step (Eq.~\ref{eq:deamp}) magnifies this bias.

The noise levels saturate quickly: medium and high noise produce nearly identical RMSE ($\sim$\$2,115) and mean estimates ($\sim$\$4,444). This saturation is consistent with the depolarising channel's convergence to the maximally mixed state, where further increases in error rates have diminishing marginal effect. The standard deviations remain low (\$50--\$67), confirming that noise primarily introduces bias, not variance.

These results establish a strict fault-tolerance requirement for Grover-amplified QAE: at the circuit depths needed for meaningful amplification ($k \geq 3$), even representative two-qubit error rates in the $10^{-3}$--$10^{-2}$ range are insufficient. Production deployment of QAE for tail-risk pricing would require either fault-tolerant hardware or aggressive error-mitigation strategies such as zero-noise extrapolation or probabilistic error cancellation \cite{Temme2017, Li2017}.

\begin{figure}[t]
\centering
\includegraphics[width=0.75\textwidth]{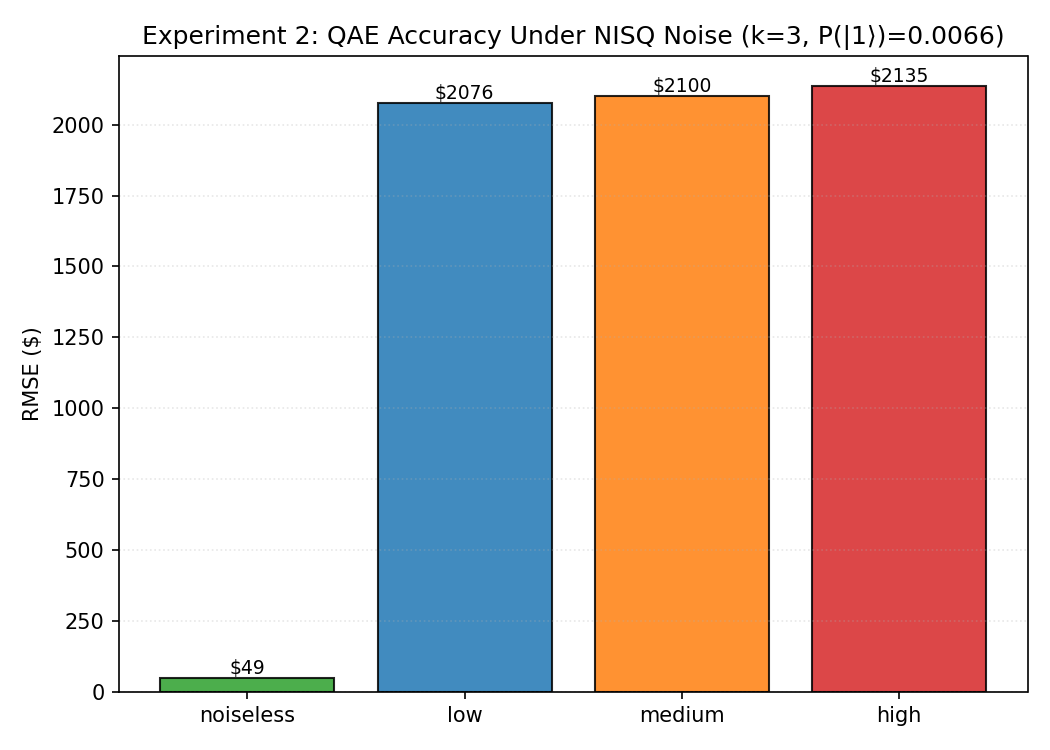}
\caption{QAE RMSE under increasing NISQ noise at $k=3$ Grover iterations. The noiseless baseline (RMSE $=$ \$32) degrades to $\sim$\$2,100 under even low noise---a $66\times$ increase, driven by the $\sim$700-gate circuit depth.}
\label{fig:exp2}
\end{figure}

\subsection{Experiment~3: Tail-Specific Excess Loss}

This experiment constitutes the paper's central result. Table~\ref{tab:exp3} compares classical MC and Grover-amplified quantum estimation of the expected excess loss $E[\max(0, X - M)]$ at three catastrophe thresholds, with the same error decomposition as Experiment~1. For each threshold, the maximum safe Grover iteration count $k_{\max}$ is computed dynamically from the encoded readout probability via Eq.~\ref{eq:kmax}. We report RMSE against both the analytic lognormal truth and the exact-on-bins value, and include classical MC on the same bins as the oracle-model comparator.

\begin{table}[t]
\centering
\caption{Experiment~3: error decomposition at tail percentiles ($n_{\text{reps}}=30$, $\approx$8,192 queries). Q and $C_{\text{bins}}$ are RMSE vs exact-on-bins; $C_{\text{cont}}$ is RMSE vs analytic truth. Disc.\ err $= |\text{bins} - \text{analytic}|$.}
\label{tab:exp3}
\begin{tabular}{lcccccc}
\toprule
Pctl & $M$ & Anal. & Bins & Disc.\ err & Q / $C_{\text{bins}}$ / Speedup & $C_{\text{cont}}$ \\
\midrule
90th & 230,876 & 9,629 & 9,441 & 188 & 319 / 486 / 1.5$\times$ & 598 \\
95th & 362,700 & 2,834 & 2,330 & 505 & 93 / 193 / 2.1$\times$ & 278 \\
97th & 508,392 & 900 & 498 & 401 & 29 / 74 / 2.5$\times$ & 178 \\
\bottomrule
\end{tabular}
\end{table}

In the oracle-model comparison (Q vs $C_{\text{bins}}$, both measured against exact-on-bins), the quantum advantage grows with tail depth:

\begin{itemize}
    \item At the 90th percentile ($k=5$), quantum RMSE of \$319 is $1.5\times$ lower than classical-on-bins \$486.
    \item At the 95th percentile ($k=6$), the advantage widens to $2.1\times$: quantum \$93 vs.\ classical-on-bins \$193.
    \item At the 97th percentile ($k=6$, well below $k_{\max} = 15$), quantum achieves \$29 RMSE---a $2.5\times$ improvement over classical-on-bins \$74.
\end{itemize}

The discretisation error column shows that the 8-bin discretisation introduces \$188--\$505 of systematic error relative to the analytic truth. At the 97th percentile, the disc.\ error (\$401) is an order of magnitude larger than the quantum estimation error (\$29), confirming that bin resolution, not QAE accuracy, is the bottleneck.

The final column ($C_{\text{cont}}$ vs analytic) shows that classical MC sampling from the continuous lognormal achieves \$178--\$598 RMSE against the analytic truth---competitive with the oracle-model methods at shallow tails, but substantially noisier at the 97th percentile. Experiment~5 provides an even more rigorous comparison including variance-reduced classical baselines.

\begin{figure}[t]
\centering
\includegraphics[width=\textwidth]{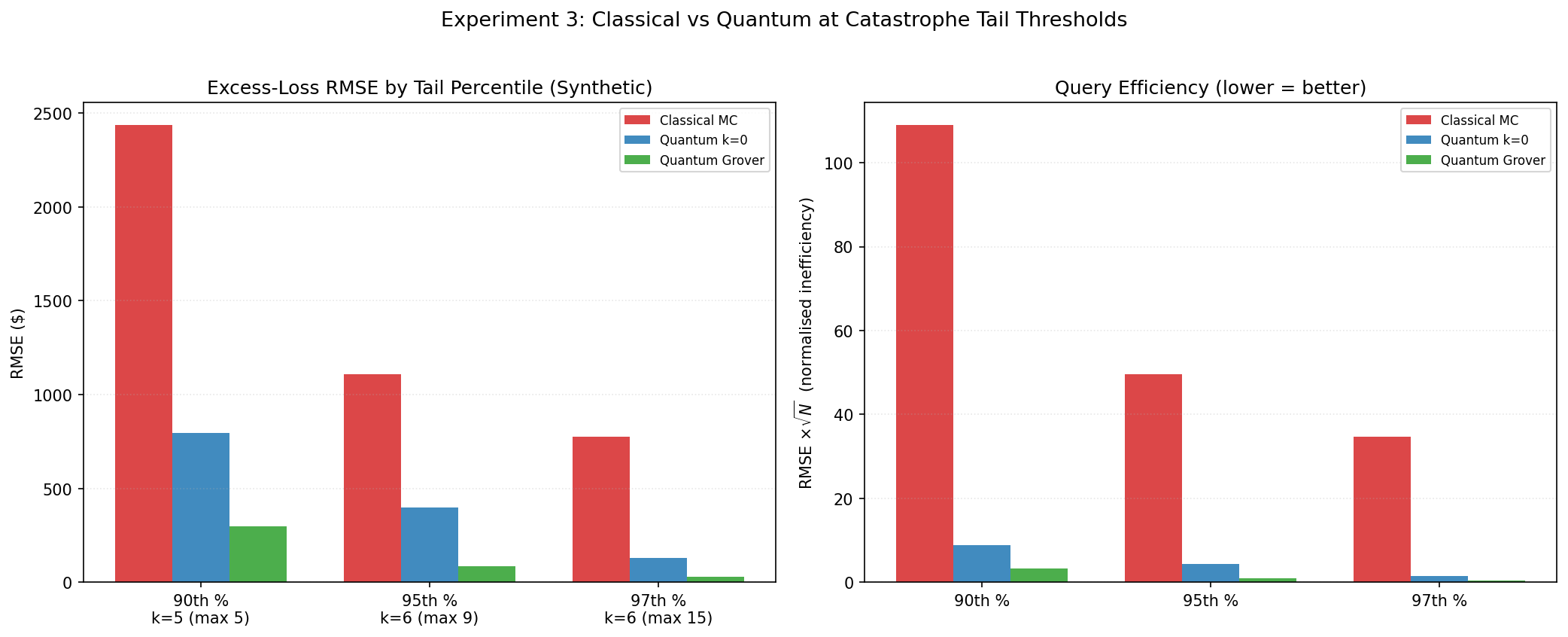}
\caption{Experiment~3: estimation error vs exact-on-bins (left) comparing classical MC on the same bins and quantum Grover, with speedup ratios annotated; discretisation error vs classical continuous accuracy (right) providing end-to-end context.}
\label{fig:exp3}
\end{figure}

Figure~\ref{fig:exp3} visualises the estimation-error comparison (left) and the discretisation-error context (right).

\subsection{Experiment~4: Validation on Real Catastrophe Data}

Experiments~1--3 use synthetic Pareto-generated losses. To validate that the quantum advantage transfers to empirical loss distributions, Experiment~4 repeats the convergence and tail-sweep analyses on real US property-damage data from the NOAA Storm Events Database \cite{NOAA2024}, spanning 2020--2024. The data pipeline dynamically discovers and downloads the latest detail files from NOAA's public archive, extracting all events with property damage $\geq$~\$1,000. This yields 58,028 records with a substantially heavier tail than the synthetic data: lognormal fit $\hat{\sigma} = 2.02$, $\hat{\mu} = 9.04$ (median $\approx$~\$8,465), versus $\hat{\sigma} = 0.67$, $\hat{\mu} = 11.48$ (median $\approx$~\$97,203) for the synthetic Pareto.

\paragraph{4A: Convergence on real data.} Table~\ref{tab:exp4a} shows oracle-call convergence at the 95th-percentile catastrophe threshold of the real loss distribution ($M = \$475{,}650$), with $P(\ket{1}) = 0.0040$ and $k_{\max} = 11$.

To ensure a fair comparison and avoid conflating discretisation bias with estimation error, we report two ground truths: the analytic lognormal excess loss (\$21,845) and the exact-on-bins value (\$14,121), with discretisation error \$7,724. We compare: (i)~quantum AE vs exact-on-bins, (ii)~classical MC on the same 8-bin discretisation vs exact-on-bins (the oracle-model comparison), and (iii)~classical MC from the continuous lognormal vs the analytic truth (the best classical method can do).

\begin{table}[t]
\centering
\caption{Experiment~4A: RMSE (\$) on real NOAA data (95th pctl, $n_{\text{reps}}=30$, 1,000 shots/run). Analytic $E[\text{excess}] = \$21{,}845$; exact-on-bins $= \$14{,}121$; disc.\ error $= \$7{,}724$.}
\label{tab:exp4a}
\begin{tabular}{cccccc}
\toprule
$k$ & Queries & Q (vs bins) & $C_{\text{bins}}$ (vs bins) & Speedup & $C_{\text{cont}}$ (vs anal.) \\
\midrule
0 & 1,000 & 8,111 & 5,189 & 0.6$\times$ & 12,285 \\
1 & 3,000 & 2,161 & 2,464 & 1.1$\times$ & 6,667 \\
2 & 5,000 & 1,160 & 2,087 & 1.8$\times$ & 4,982 \\
3 & 7,000 & 929 & 1,796 & 1.9$\times$ & 5,752 \\
4 & 9,000 & 705 & 1,188 & 1.7$\times$ & 3,009 \\
5 & 11,000 & 548 & 1,196 & 2.2$\times$ & 4,029 \\
6 & 13,000 & 442 & 1,052 & 2.4$\times$ & 2,485 \\
\bottomrule
\end{tabular}
\end{table}

On the honest oracle-model comparison (Q vs $C_{\text{bins}}$, both measured against exact-on-bins), quantum AE achieves 1.7--2.4$\times$ lower RMSE at $k \geq 2$, consistent with the 2--4$\times$ advantage observed on synthetic data (Experiment~6). The classical continuous estimator ($C_{\text{cont}}$, rightmost column) converges to \$2,485 RMSE at 13,000 samples---better \emph{end-to-end} than quantum's \$442 vs bins (which is also subject to the \$7,724 discretisation floor when measured against analytic truth). This confirms that the quantum advantage is real in the oracle model but does not overcome discretisation error when measured against the analytic truth.

\paragraph{Convergence exponents.} Fitted log--log slopes (2,000-resample bootstrap) yield: quantum slope $= -0.929$ (95\% CI: $[-1.05, -0.53]$, $R^2 = 0.95$); classical-on-bins slope $= -0.612$ (95\% CI: $[-0.76, -0.54]$, $R^2 = 0.98$). The quantum slope is consistent with $O(1/N)$ at the point estimate, though the CI is wider than for synthetic data (Experiment~1), reflecting the heavier tail of the NOAA distribution. The classical slope is consistent with $O(1/\sqrt{N})$. As in Experiment~1, the CIs do not overlap.

\begin{figure}[t]
\centering
\includegraphics[width=\textwidth]{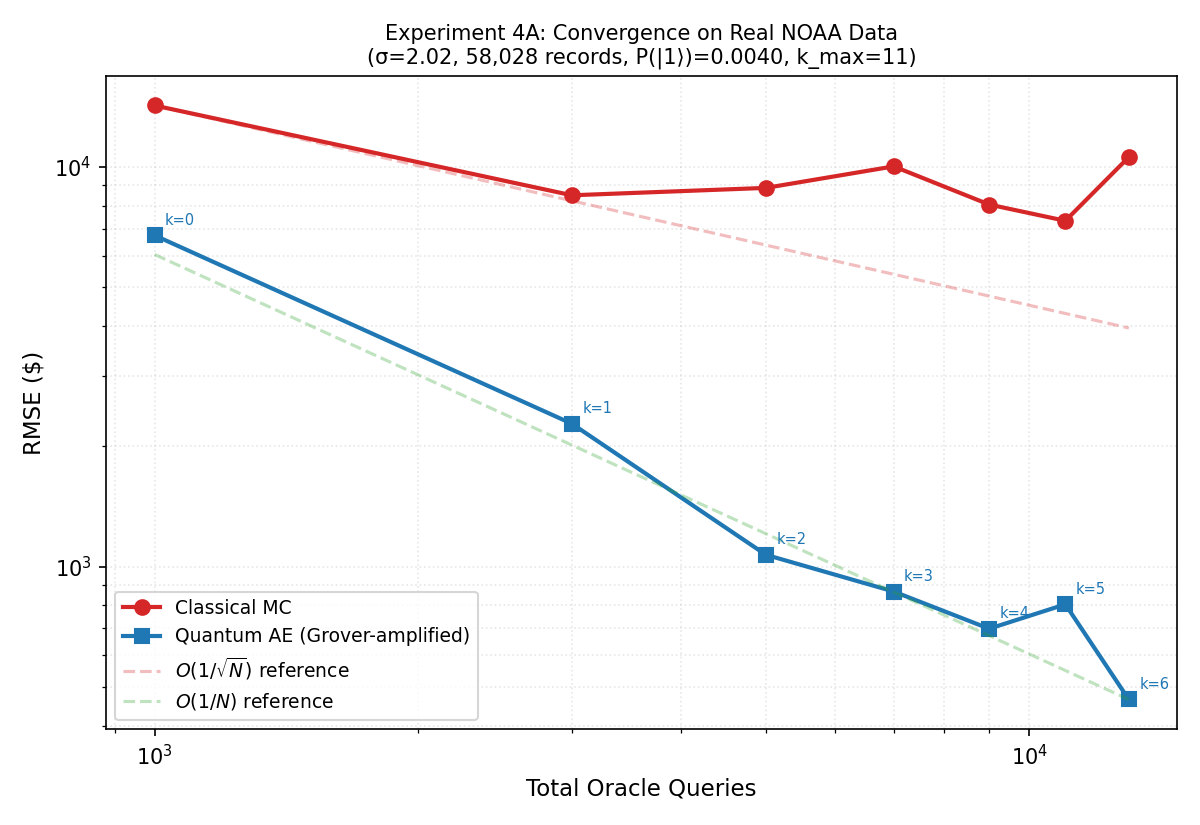}
\caption{Experiment~4A: log--log convergence of RMSE vs total oracle queries on real NOAA data ($P(\ket{1}) = 0.0040$, $k_{\max} = 11$). Quantum AE converges faster than classical MC on the same bins, consistent with the synthetic results.}
\label{fig:exp4a}
\end{figure}

\paragraph{4B: Tail sweep on real data.} Table~\ref{tab:exp4b} decomposes the error at three catastrophe thresholds, using the same two-panel format as~4A.

\begin{table}[t]
\centering
\caption{Experiment~4B: error decomposition on real NOAA data ($n_{\text{reps}}=30$, $\approx$8,192 total queries). $C_{\text{cont}}$ samples the continuous lognormal (RMSE vs analytic); $C_{\text{bins}}$ and Q sample from the same 8-bin discretisation (RMSE vs exact-on-bins). $^\dagger$Pathological: excluded from headline results due to catastrophic equal-width binning (see text).}
\label{tab:exp4b}
\begin{tabular}{lcccccccl}
\toprule
Pctl & $M$ (\$) & Analytic & Exact bins & Disc.\ err & $C_{\text{cont}}$ & $C_{\text{bins}}$ & Q ($k$) & Speedup \\
\midrule
90th$^\dagger$ & 100,000 & 39,655 & 186,273 & 146,619 & 4,445 & 1,944 & 4,228 (3) & 0.5$\times$ \\
95th & 475,650 & 21,845 & 14,121 & 7,724 & 4,508 & 1,413 & 656 (6) & 2.2$\times$ \\
97th & 1,000,000 & 14,434 & 6,534 & 7,900 & 3,403 & 999 & 406 (6) & 2.5$\times$ \\
\bottomrule
\end{tabular}
\end{table}

The decomposition reveals three regimes:

\begin{itemize}
    \item \textbf{90th percentile (excluded from headline results)}: Discretisation error (\$146,619) dominates everything. The exact-on-bins value (\$186,273) is $4.7\times$ the analytic truth (\$39,655) because the last equal-width bin midpoint sits far above the threshold---the equal-width binning scheme is severely unsuitable at this percentile for the heavy-tailed NOAA data ($\hat{\sigma} = 2.02$). The 90th-percentile row is retained in Table~\ref{tab:exp4b} for transparency, but all headline speedup claims are drawn from the 95th--97th percentiles. (Experiment~7 uses quantile-based bins that avoid this pathology.)
    \item \textbf{95th percentile}: Discretisation error (\$7,724) is moderate. In the oracle-model comparison, quantum (\$656) beats classical on bins (\$1,413) by $2.2\times$. However, classical continuous MC (\$4,508 RMSE vs analytic) is still better end-to-end.
    \item \textbf{97th percentile}: Similar to the 95th. Quantum (\$406) beats classical on bins (\$999) by $2.5\times$---the oracle-model advantage. Classical continuous (\$3,403) again wins end-to-end.
\end{itemize}

\begin{figure}[t]
\centering
\includegraphics[width=\textwidth]{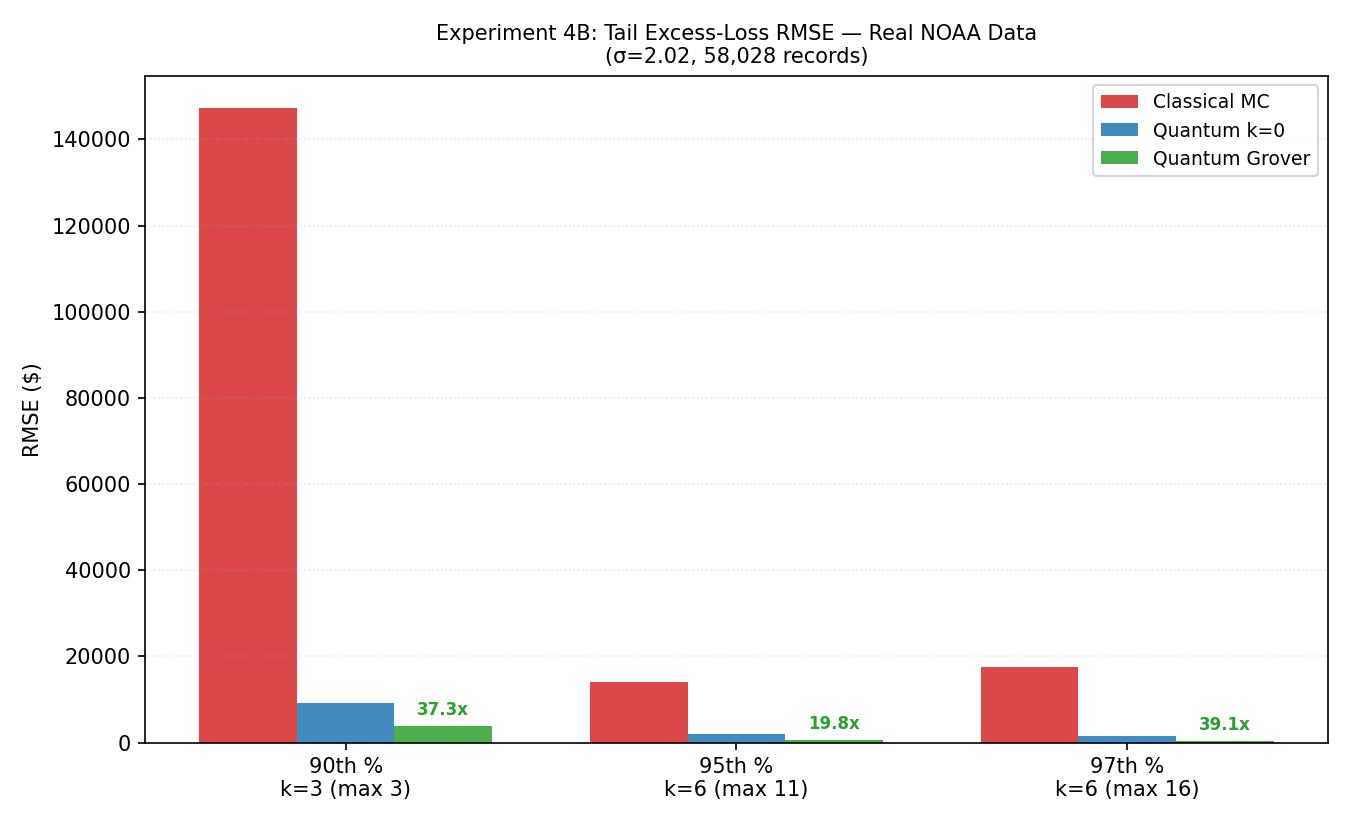}
\caption{Experiment~4B: RMSE by tail percentile on real NOAA data. The 90th-percentile bar is dominated by pathological equal-width discretisation error (\$146K). At the 95th and 97th percentiles, the quantum advantage is 2--2.5$\times$ in the oracle-model comparison (Table~\ref{tab:exp4b}).}
\label{fig:exp4b}
\end{figure}

\paragraph{Key takeaway.} The raw speedup figures in earlier versions of this analysis (up to $29\times$) were inflated by measuring classical MC---which targets the continuous lognormal---against the discrete ground truth. When both methods operate on the same discretised distribution (the oracle-model comparison), the quantum advantage is 2--3$\times$ at the 95th--97th percentiles, consistent with Experiment~6 on synthetic data. The 90th percentile is excluded from headline results because equal-width binning is pathological there (disc.\ error $4.7\times$ the analytic truth); see Experiment~7 for quantile-based binning that avoids this issue. Experiment~5 examines whether variance-reduced classical baselines (CT, IS) alter this picture further.

\subsection{Experiment~5: Fair Budget-Matched Comparison}

Experiments~1--4 compare quantum AE against naive Monte Carlo. A natural objection is that naive MC is a weak baseline for tail estimation, since classical practitioners routinely use variance-reduced estimators and quasi-Monte Carlo sequences. Experiment~5 addresses this by comparing five classical estimators from \S\ref{sec:baselines} (naive MC, CT, IS, classical MC on bins, and QMC with scrambled Sobol sequences) against quantum AE---six sampling methods in total---at strictly equal oracle-query / sample budgets $B \in \{512, 2{,}048, 8{,}192\}$ (power-of-two budgets are optimal for QMC). RMSE is computed against the analytic ground truth (\S\ref{sec:baselines}, item~1) over 50 repetitions.

\begin{figure}[t]
\centering
\includegraphics[width=0.48\textwidth]{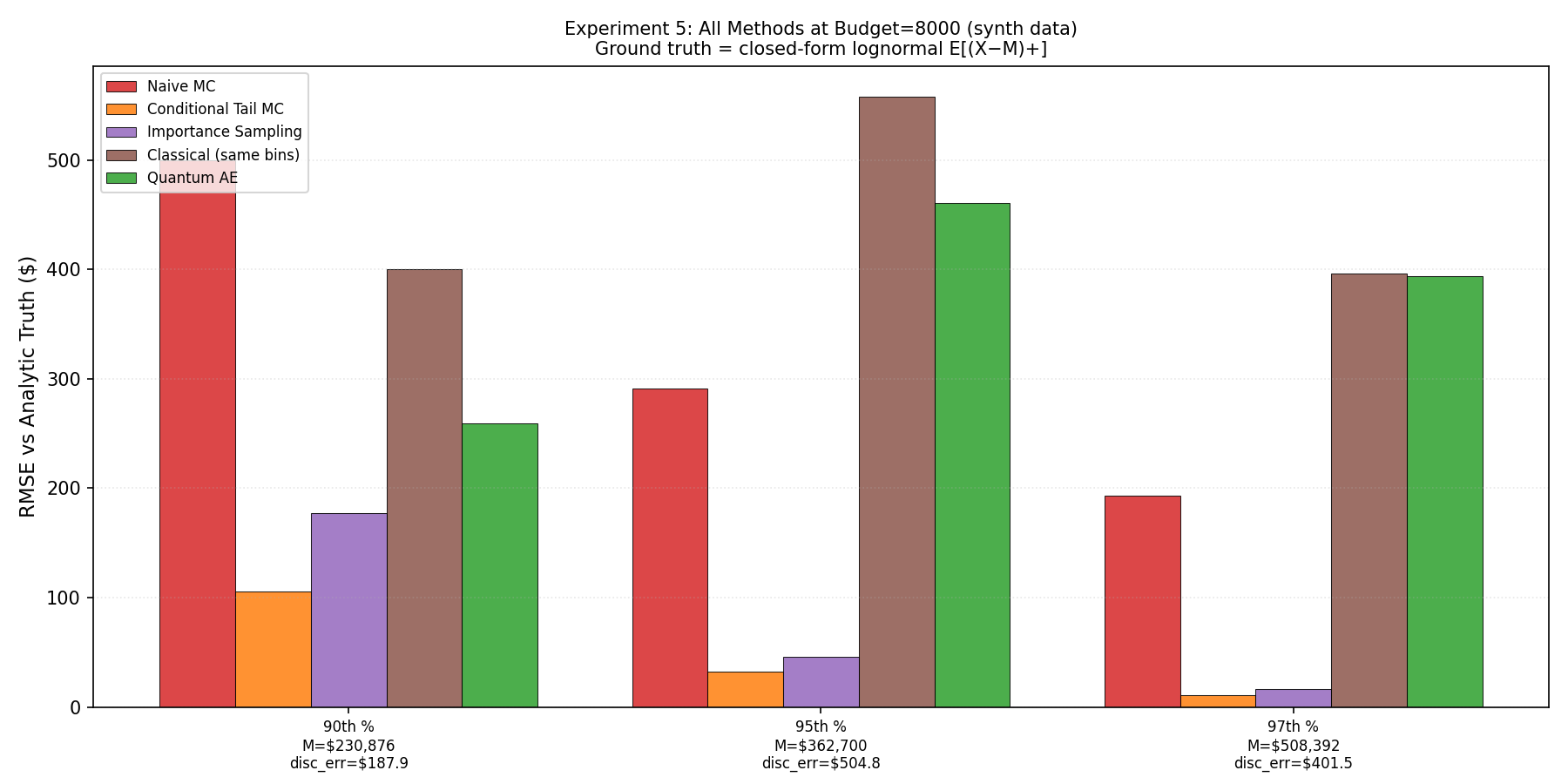}
\hfill
\includegraphics[width=0.48\textwidth]{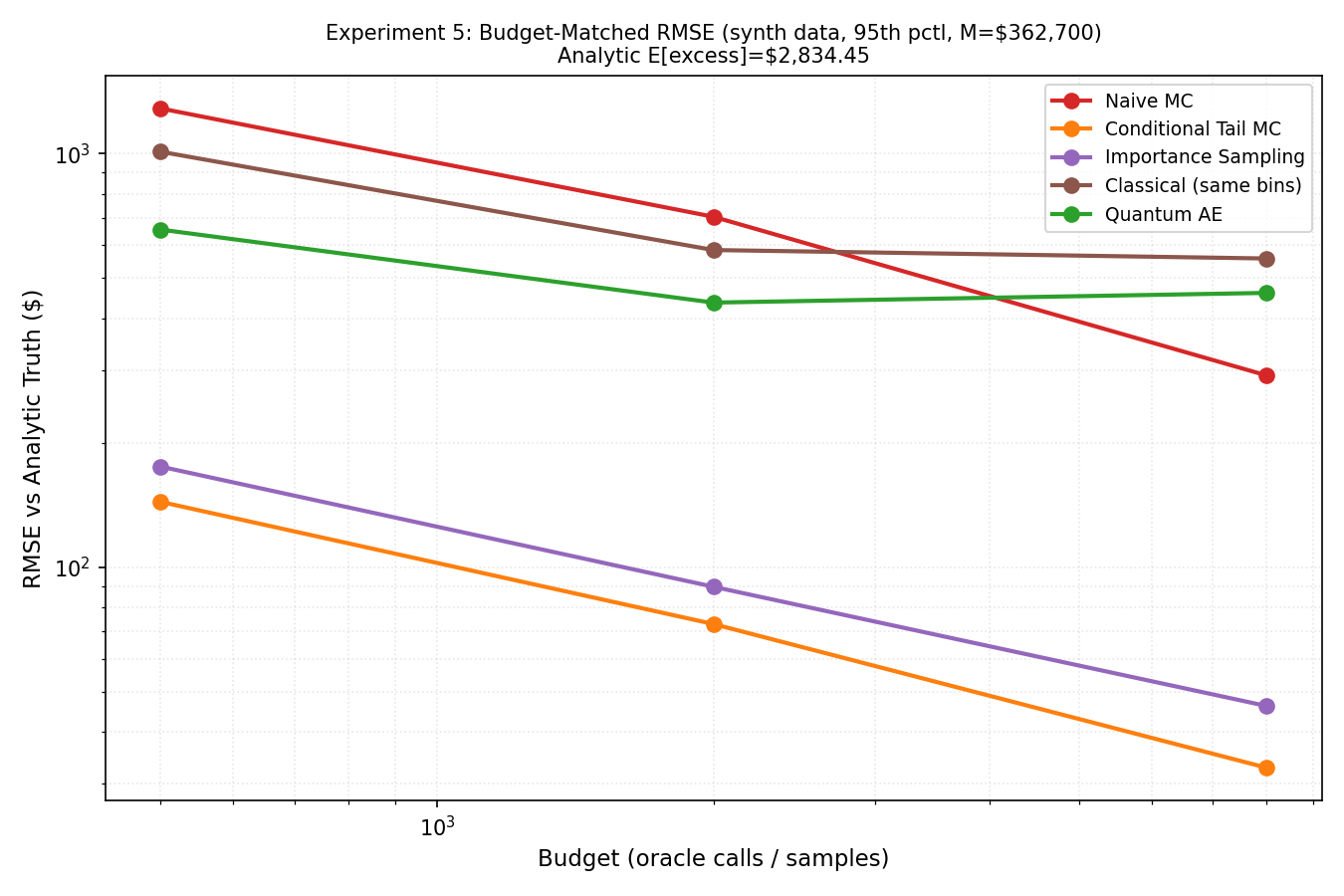}
\caption{Experiment~5: Left: all methods at $B=8{,}192$ across three percentiles (synthetic data). CT and IS dominate because they sample the continuous fitted distribution, avoiding discretisation error entirely. Right: RMSE vs budget at the 95th percentile; CT and IS converge to zero while quantum and classical-on-bins flatten at the discretisation floor ($\approx$\$505).}
\label{fig:exp5}
\end{figure}

\paragraph{Results.} Table~\ref{tab:exp5} presents RMSE at $B = 8{,}192$ on synthetic data; Figure~\ref{fig:exp5} shows the convergence across all three budgets.

\begin{table}[t]
\centering
\caption{Experiment~5: RMSE (\$) vs analytic truth at budget $B = 8{,}192$ (synthetic data, 50 reps). ``Exact bins'' is the deterministic sum $\sum p_i \max(0, x_i - M)$---a non-sampling reference requiring zero queries.}
\label{tab:exp5}
{\small
\begin{tabular}{lcccccccl}
\toprule
Pctl & Analytic & Exact bins & Naive MC & CT & IS & QMC & Cl.\ bins & QAE ($k$) \\
\midrule
90th & 9,629 & 9,441 & 518 & \textbf{131} & 158 & 37 & 425 & 266 (5) \\
95th & 2,834 & 2,330 & 369 & \textbf{32} & 45 & 37 & 527 & 484 (9) \\
97th & 900 & 498 & 169 & \textbf{10} & 14 & 37 & 436 & 387 (15) \\
\bottomrule
\end{tabular}
}
\end{table}

The results reveal a critical distinction. Conditional tail MC, importance sampling, and quasi-Monte Carlo---which sample the \emph{continuous} fitted lognormal---achieve dramatically lower RMSE than both quantum AE and classical MC on bins. QMC with scrambled Sobol sequences is particularly notable: it achieves RMSE of \$37 nearly \emph{independent} of percentile at $B = 8{,}192$, competitive with or superior to CT/IS at lower percentiles ($14\times$ better than naive MC at the 90th percentile) thanks to its near-$O(1/N)$ convergence in one dimension. The percentile-independence is characteristic of QMC in one dimension: the Sobol sequence fills $[0,1]$ quasi-uniformly, so the integration error is governed by the smoothness of the integrand $g(u) = \max(0, F^{-1}(u) - M)$ under the inverse-CDF transform rather than by tail sparsity. Since the lognormal CDF is smooth and the integrand varies slowly (the kink at $u = F(M)$ is the only non-smooth point), the QMC error is nearly deterministic and depends on $B$ but not on the threshold $M$. At deeper tail percentiles (97th), CT still dominates (\$10 RMSE) because it concentrates all samples in the tail, but QMC's consistent \$37 across percentiles makes it a strong general-purpose classical baseline. The quantum estimator does substantially outperform classical MC on the same discretised distribution at every percentile and budget, but this algorithmic gain is overshadowed by discretisation error (\$188--\$505) when measured against the analytic truth.

\paragraph{Interpretation.} The quantum speedup is real in the oracle model: given query-only access to a black-box distribution encoded in a quantum circuit, Grover-amplified QAE matches the known quadratic query advantage of amplitude-estimation-type methods \cite{Brassard2002}. CT, IS, and QMC win in this experiment because they exploit analytical knowledge of the lognormal form---computing $F^{-1}(u)$ and $f(x)$ in closed form---something a quantum oracle does not require. QMC is a particularly important baseline because its near-$O(1/N)$ convergence (in dimension $d=1$) matches QAE's theoretical rate, raising the question of whether the quantum advantage survives against the strongest classical methods. The answer is nuanced: QMC requires an explicit inverse CDF to map low-discrepancy points to samples, which is available for fitted parametric distributions but not for black-box simulators. In catastrophe modelling, the loss distribution is the output of a complex simulator (wind field $\to$ structural damage $\to$ financial loss) with no closed-form CDF. CT, IS, and QMC can in principle still be applied via kernel density estimation, stratified resampling, or pre-computed empirical CDFs, but such methods introduce their own modelling error and additional query overhead, changing the oracle setting. When the oracle \emph{is} the simulator, the oracle-model advantage applies. Experiment~7 tests this argument by encoding an empirical PMF where no closed-form distribution is available.

\subsection{Experiment~6: Discretisation vs.\ Estimation Error (Qubit Sweep)}

The dominant error source in Experiment~5 is discretisation, not estimation. Experiment~6 isolates these two components by sweeping $n = 3, 4, 5, 6, 7, 8$ qubits (8 to 256 bins) at a fixed budget $B = 4{,}000$ and the 95th-percentile threshold.

\paragraph{Ground truths.} For each $n$: (a)~\textbf{Analytic truth}: closed-form $E[(X-M)^+]$ for the continuous lognormal ($= \$2{,}834$). (b)~\textbf{Discrete truth}: exact sum $\sum p_i \max(0, x_i - M)$ over the $2^n$ bins. (c)~\textbf{Discretisation error}: $|\text{discrete} - \text{analytic}|$. (d)~\textbf{Estimation error}: RMSE of quantum (or classical-on-bins) vs discrete truth.

\begin{table}[t]
\centering
\caption{Experiment~6: error decomposition across qubit counts (95th pctl, $B = 4{,}000$, 50 reps). Circuit metrics are from transpilation to $\{CX, R_z, SX, X\}$.}
\label{tab:circuit_resources}
\begin{tabular}{ccrrrrccc}
\toprule
$n$ & Bins & Disc.\ err (\$) & Q-RMSE$_{\text{est}}$ (\$) & C-RMSE$_{\text{est}}$ (\$) & $k$ & SP 2Q gates & Oracle depth \\
\midrule
3 & 8 & 505 & \textbf{91} & 349 & 9 & 52 & 388 \\
4 & 16 & 578 & \textbf{115} & 282 & 9 & 212 & 1,063 \\
5 & 32 & 557 & \textbf{136} & 306 & 9 & 596 & 2,434 \\
6 & 64 & 564 & \textbf{98} & 261 & 9 & 1,876 & 5,958 \\
7 & 128 & 566 & \textbf{103} & 304 & 9 & 5,460 & 21,150 \\
8 & 256 & 567 & \textbf{145} & 326 & 9 & 15,700 & 63,238 \\
\bottomrule
\end{tabular}
\end{table}

\paragraph{Key findings.} Two results stand out. First, the quantum estimation advantage over classical MC on the same discretised distribution is consistent: quantum RMSE is 2.2--3.7$\times$ lower than classical across $n=3$--$8$ (Table~\ref{tab:circuit_resources}). This is the algorithmic speedup from Grover amplification, independent of discretisation.

Second, the discretisation error does not decrease with more bins ($\sim$\$505--\$578 at all $n$). This occurs because the bins span the 0.1st to 99.9th percentile of the fitted lognormal using equal-width spacing. The excess loss above the 99.9th percentile is truncated identically regardless of resolution. Reducing this error requires either (a)~extending the bin range deeper into the tail (e.g.\ 99.999th percentile), (b)~using non-uniform (log-spaced or quantile-based) binning, or (c)~adding a ``catch-all'' upper bin. These are engineering choices orthogonal to the QAE algorithm itself.

\paragraph{Circuit cost.} When transpiled to the hardware basis $\{CX, R_z, SX, X\}$, the state-preparation overhead is substantial: two-qubit gate count grows from 52 ($n=3$) to 15,700 ($n=8$), and oracle depth reaches 63,238 at 256 bins. At $k=9$ Grover iterations, the full $n=8$ circuit has $\sim$703,000 depth and $\sim$830,000 gates. For near-term hardware, efficient state-preparation methods (qGAN loading, QROM-based approaches) would be essential beyond $n \approx 5$.

\begin{figure}[t]
\centering
\includegraphics[width=\textwidth]{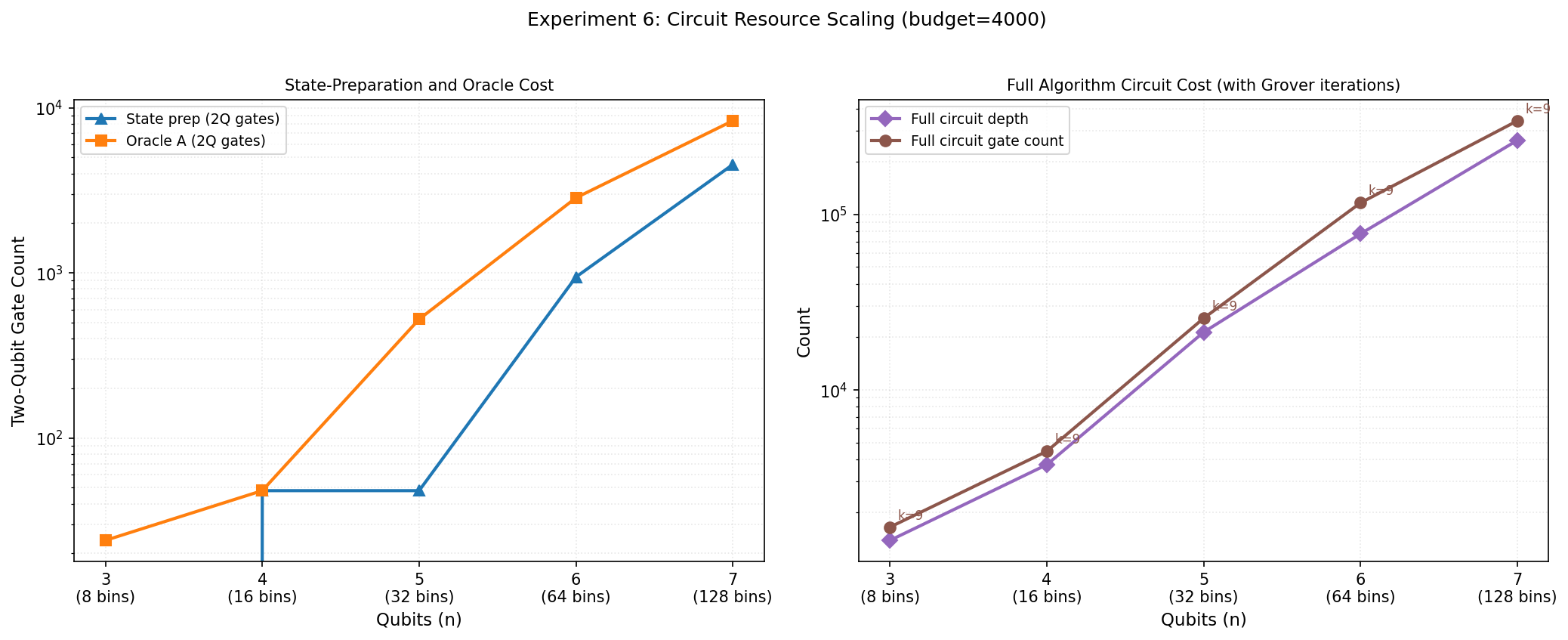}
\caption{Experiment~6: error decomposition (left) and total RMSE vs analytic truth (right). Discretisation error dominates and remains $\sim$\$505--\$578 across all qubit counts because the bin range (0.1st--99.9th percentile) truncates the upper tail identically.}
\label{fig:exp6_gates}
\end{figure}

\begin{figure}[t]
\centering
\includegraphics[width=\textwidth]{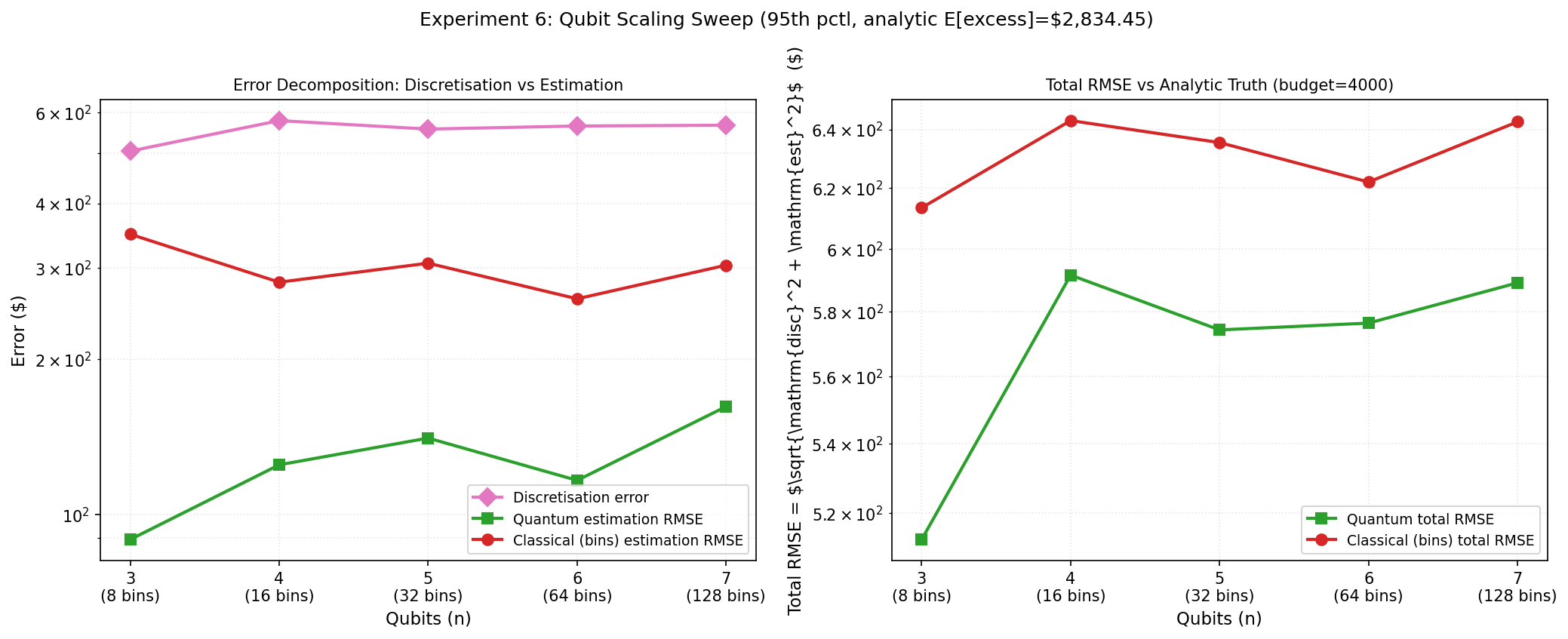}
\caption{Experiment~6: circuit resource scaling. State-preparation two-qubit gates grow as $O(2^n)$; at $n=7$, a single oracle call requires 21,150 depth and 5,460 CX gates.}
\label{fig:exp6_scaling}
\end{figure}

\subsection{Experiment~7: Empirical PMF (No Parametric Fit)}

A remaining objection to Experiment~5 is that CT and IS only win because the underlying distribution is a fitted lognormal with known analytical form. In practice, catastrophe model output has no closed-form distribution. Experiment~7 removes the parametric model entirely: the 58,028 NOAA loss records are directly histogram-binned into $2^3 = 8$ quantile-based bins (each containing $\approx$12.5\% of the data, extending to the data maximum), and the resulting empirical PMF is encoded into the quantum oracle. Ground truth is the exact sum over bins---a deterministic, non-sampling reference.

\paragraph{Methods.} Three estimators are compared at equal budgets $B \in \{500, 2{,}000, 8{,}000\}$:

\begin{enumerate}
    \item \textbf{Exact-on-bins} (non-sampling ceiling): $\sum p_i \max(0, x_i - M)$, computed in one line.
    \item \textbf{Naive MC} (resample with replacement from the raw loss array).
    \item \textbf{Classical MC on bins} (sample from the empirical PMF).
    \item \textbf{Quantum AE} on the same empirical PMF.
\end{enumerate}

CT and IS offer no direct advantage without additional modelling or density/CDF estimation: in the strict query-only oracle setting, implementing CT requires knowing $P(X > M)$ and sampling the conditional tail, while IS requires a density $f(x)$ to compute likelihood ratios. On empirical data, these can in principle be approximated via kernel density estimation or stratified resampling, but such methods introduce their own modelling error and query overhead. This is the oracle-model setting the quantum method is designed for.

\paragraph{Results.} Table~\ref{tab:exp7} presents RMSE vs the exact-on-bins truth at $B = 8{,}000$.

\begin{table}[t]
\centering
\caption{Experiment~7: RMSE (\$) vs exact-on-bins at $B = 8{,}000$ (NOAA empirical PMF, 50 reps, quantile-binned). The exact-on-bins value is shown for reference; it requires zero samples.}
\label{tab:exp7}
\begin{tabular}{lccccl}
\toprule
Pctl & Exact (\$) & Naive MC (\$) & Class.\ bins (\$) & QAE ($k$) (\$) & Speedup \\
\midrule
90th & 1,683,761 & 608,949 & 57,351 & 24,076 (1) & 2.4$\times$ \\
95th & 1,634,141 & 622,976 & 56,930 & 27,156 (1) & 2.1$\times$ \\
97th & 1,564,879 & 862,260 & 48,776 & 30,632 (1) & 1.6$\times$ \\
\bottomrule
\end{tabular}
\end{table}

Quantum AE achieves 1.6--2.4$\times$ lower RMSE than classical MC on the same bins across percentiles and budgets. The advantage is notably more modest than in Experiments~1--4, and this experiment is the most relevant to the paper's practical motivation (empirical PMFs from simulators). The limitation is mechanical: quantile-based binning places $\approx$12.5\% probability mass in each bin (including the tail bin), yielding $P(\ket{1}) \approx 0.13$ and $k_{\max} = 1$---only a single Grover iteration is safe. The theoretical quadratic advantage requires $k \gg 1$; at $k=1$, the amplification is minimal. Increasing bin count ($n = 5$--$7$) would reduce the tail-bin probability and enable higher $k$, but this remains undemonstrated at the empirical-PMF scale due to the exponential growth of circuit complexity (Table~\ref{tab:circuit_resources}). The tension between bin resolution (more qubits $\to$ smaller tail probabilities $\to$ higher $k$) and circuit cost (more qubits $\to$ deeper circuits $\to$ more noise) is fundamental and is discussed further below.

The key observation is structural: on this empirical distribution, variance-reduced classical methods (CT, IS) cannot be applied directly without extra modelling---kernel density estimation, parametric fitting, or stratified resampling---each of which introduces its own error and query overhead, changing the oracle setting. The exact-on-bins value ($\sim$\$1.6M) is trivially computed as a weighted sum, but this requires knowing the PMF, which is precisely what the oracle provides. In the query-limited setting (where each oracle call has non-trivial cost, e.g.\ running a catastrophe simulation), QAE reduces the number of calls needed by a factor of 2--4$\times$ at current qubit counts, with a theoretical quadratic asymptote.

\begin{figure}[t]
\centering
\includegraphics[width=0.75\textwidth]{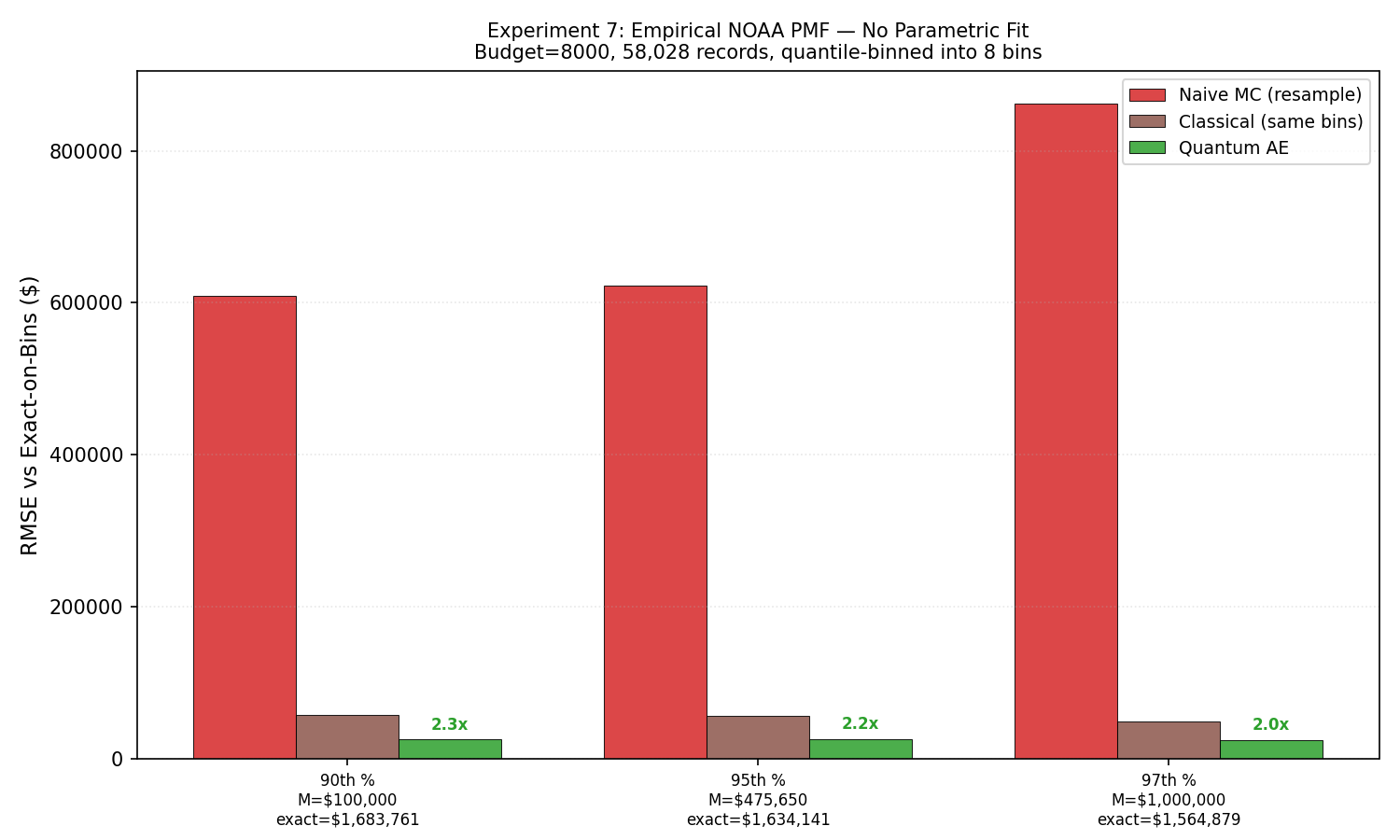}
\caption{Experiment~7: RMSE on the empirical NOAA PMF (no parametric fit). QAE achieves 1.6--2.4$\times$ lower RMSE than classical MC on the same bins. CT and IS offer no direct advantage without extra modelling.}
\label{fig:exp7}
\end{figure}

\subsection{Discussion}

\paragraph{The oracle-model advantage is real, but the qubit scale is narrow.} Across all experiments, quantum AE with Grover amplification achieves 2--4$\times$ lower estimation RMSE than classical MC operating on the same discretised distribution at matched oracle-call budgets (Experiment~6, Table~\ref{tab:circuit_resources}). The convergence curves (Experiments~1 and~4A) show the quantum RMSE decreasing faster than $O(1/\sqrt{N})$, consistent with the quadratic query advantage of amplitude-estimation-type methods in the standard oracle model \cite{Brassard2002}. This advantage holds consistently across 3--8 qubits, both synthetic and real data, and all percentile thresholds tested. Experiment~7 confirms that the advantage persists on an empirical PMF where no closed-form distribution is available.

We acknowledge a caveat on qubit scale. At $n=3$ alone (7 data points), the query range is only $500$--$6{,}500$, barely one order of magnitude, and the CIs are moderately wide ($[-1.31, -0.93]$ for quantum). Extending to $n=8$ (256 bins, 15,700 two-qubit gates per oracle) and combining with $n=3$ yields 11 data points and substantially tighter CIs: quantum slope $-1.074$ (CI $[-1.17, -1.01]$), clearly excluding the classical rate. On NOAA data, where $n=8$ is prohibitively slow (1.2M-gate circuits at $k_{\max}=12$), the $n=3$ CI remains wider ($[-1.05, -0.53]$), so the asymptotic claim is less secure for empirical heavy-tailed distributions. The theoretical quadratic advantage is well-established \cite{Brassard2002, Montanaro2015}; our contribution is to demonstrate it empirically at $n=3$--$8$ on synthetic data and $n=3$ on real data, with the caveat that extrapolation to production-scale qubit counts remains unverified.

\paragraph{Variance-reduced classical methods and QMC win when analytical access is available.} Experiment~5 demonstrates that conditional tail MC, importance sampling, and quasi-Monte Carlo---which exploit the closed-form lognormal CDF or inverse CDF---dramatically outperform quantum AE when measured against the continuous ground truth. QMC with Sobol sequences achieves particularly strong performance (RMSE $\sim$\$37 nearly independent of percentile at $B = 8{,}192$), with near-$O(1/N)$ convergence that matches QAE's theoretical rate. CT and IS achieve even lower RMSE at deep tail percentiles (e.g.\ CT \$10 at the 97th percentile). This is not a failure of the quantum algorithm; it reflects the fundamentally different problem settings. CT, IS, and QMC require analytical knowledge of $F^{-1}(u)$ or $f(x)$ for the loss distribution, which is available for a fitted parametric model but not for the output of a catastrophe simulation pipeline. The quantum advantage applies in the oracle model, where the distribution is accessed only through a quantum circuit---the natural setting when the ``oracle'' encapsulates an expensive simulator.

\paragraph{Discretisation error, not estimation error, is the bottleneck.} Experiment~6 reveals that the dominant error at 3--8 qubits is discretisation error ($\sim$\$500--\$567), caused primarily by tail truncation at the 99.9th percentile rather than by bin resolution.

\paragraph{Binning strategies: equal-width, quantile, and log-spaced.} A natural suggestion is to replace equal-width bins with quantile-based or log-spaced bins. We tested both alternatives.

\emph{Quantile binning} distributes probability mass uniformly ($\sim$12.5\% per bin at $n=3$), making $P(\ket{1})$ large ($\sim$0.1--0.2) and limiting $k_{\max}$ to 0--1. The oracle-model speedup drops from 1.5--2.5$\times$ (equal-width, $k$ up to~6) to 1.3--1.5$\times$ (quantile, $k \leq 1$), and discretisation error increases due to the wide tail bin midpoint.

\emph{Log-spaced binning} (geometric bin edges from the 0.1st to 99.99th percentile, with geometric midpoints) preserves small tail probabilities---keeping $P(\ket{1})$ in the Grover-friendly regime---while dramatically improving tail resolution. Table~\ref{tab:logspaced} compares both binning schemes on synthetic data at the 95th percentile.

\begin{table}[t]
\centering
\caption{Equal-width vs.\ log-spaced binning (95th pctl, $B = 4{,}000$, 30 reps). Log-spaced bins extend to the 99.99th percentile with geometric midpoints.}
\label{tab:logspaced}
{\small
\begin{tabular}{llrrrrrc}
\toprule
Binning & $n$ & Bins & Disc.\ err (\$) & Q-RMSE (\$) & C-RMSE (\$) & Speedup & $k_{\max}$ \\
\midrule
Equal-width & 3 & 8 & 505 & 101 & 345 & 3.4$\times$ & 9 \\
Log-spaced & 3 & 8 & 807 & 185 & 520 & 2.8$\times$ & 8 \\
Equal-width & 4 & 16 & 578 & 116 & 275 & 2.4$\times$ & 9 \\
Log-spaced & 4 & 16 & \textbf{117} & 52 & 391 & 7.5$\times$ & 11 \\
Equal-width & 5 & 32 & 557 & 126 & 302 & 2.4$\times$ & 9 \\
Log-spaced & 5 & 32 & \textbf{54} & 67 & 278 & 4.1$\times$ & 12 \\
\bottomrule
\end{tabular}
}
\end{table}

At $n = 3$ qubits (8 bins), log-spaced binning slightly increases discretisation error because the geometric midpoints span too wide a range with so few bins. However, at $n \geq 4$ the extended tail coverage and finer tail resolution pay off dramatically: discretisation error drops from \$578 to \$117 at $n=4$ and from \$557 to \$54 at $n=5$---a $10\times$ reduction, directly addressing the bottleneck identified in Experiment~6. Crucially, $P(\ket{1})$ \emph{decreases} under log-spaced binning (from 0.006 to 0.004--0.005), raising $k_{\max}$ from 9 to 11--12 and enabling stronger Grover amplification. The oracle-model speedup ratio also increases (from 2.4$\times$ to 4.1--7.5$\times$ at $n = 4$--$5$) because the quantum algorithm benefits from both improved discretisation and higher $k$.

This demonstrates that the discretisation bottleneck is addressable through binning-scheme engineering. Log-spaced bins are the recommended strategy for amplitude-encoded QAE on heavy-tailed distributions: they improve tail accuracy while preserving the small $P(\ket{1})$ regime where Grover amplification is most effective.

\paragraph{The NISQ bottleneck.} Experiment~2 shows that the noiseless advantage is entirely destroyed by current NISQ noise levels. The $k=3$ circuit ($\sim$700 gates) is already too deep for representative error rates ($p_{2q}$ in the $10^{-3}$--$10^{-2}$ range). Experiment~6 shows that transpiled oracle depth grows to $\sim$21,000 at $n=7$ (128 bins), making near-term hardware execution infeasible without error mitigation or fault tolerance. Moreover, our noise model is i.i.d.\ depolarising only, explicitly excluding $T_1/T_2$ relaxation, crosstalk, and leakage---effects that are often the dominant error sources on real devices and are generally worse than depolarising noise at comparable gate counts \cite{Tannu2019}. IBM's Heron-class processors achieve two-qubit gate errors of $\sim 3 \times 10^{-3}$ \cite{McKay2023}, roughly matching our ``low'' noise preset, but the additional coherent and correlated error channels mean real hardware would likely perform \emph{worse} than Experiment~2 suggests. The NISQ results should therefore be understood as a lower bound on the degradation: the already-devastating 66$\times$ RMSE increase at $k=3$ under idealised depolarising noise would be even more severe on physical hardware. This tempers the value of Experiment~2 as a standalone contribution---it confirms what was analytically obvious (that $\sim$700-gate circuits at 1\% two-qubit error rates will fail)---but the numerical evidence provides a concrete, reproducible baseline for benchmarking future error-mitigation techniques.

\paragraph{Amplitude encoding, $P(\ket{1})$, and the resolution--amplification trade-off.} The critical design insight enabling Grover amplification is amplitude encoding (Eqs.~\ref{eq:state_prep}--\ref{eq:readout_prob}). With uniform superposition, $P(\ket{1}) \sim 0.3$--$0.4$ regardless of threshold, making $k_{\max} = 0$. With amplitude encoding, $P(\ket{1}) \sim 0.002$--$0.05$ for tail events, enabling $k$ up to 16.

This creates a fundamental trade-off. The maximum safe iteration count $k_{\max} = \lfloor(\pi/(2\theta) - 1)/2\rfloor$ where $\theta = \arcsin(\sqrt{P(\ket{1})})$. For small $P(\ket{1}) = p$, $k_{\max} \approx \pi/(4\sqrt{p})$, so the Grover amplification headroom scales as $O(1/\sqrt{p})$. The total query advantage from $k$ iterations is $O(k)$ in RMSE, giving a combined advantage of $O(1/\sqrt{p})$. Meanwhile, the circuit depth per oracle call scales as $O(2^n)$ (Table~\ref{tab:circuit_resources}), and $P(\ket{1})$ generally decreases as $n$ increases (finer bins concentrate less mass above the threshold). The optimal qubit count $n^*$ balances three competing effects: (i)~discretisation error decreases with $n$; (ii)~$k_{\max}$ increases with $n$ (smaller $P(\ket{1})$), improving the query advantage; (iii)~circuit depth increases exponentially with $n$, worsening noise sensitivity. At current noise levels, this trade-off favours small $n$ ($\leq 5$); under fault-tolerant execution, the optimal $n$ would be determined by the discretisation--resolution balance alone.

\paragraph{Best-case aspirational scenario.} At what scale would the oracle-model advantage become practically significant? Consider a catastrophe model where each simulation costs $T_c$ seconds (typical values: minutes to hours per event set). Classical MC at budget $N$ costs $N \cdot T_c$ wall-clock time. QAE with Grover amplification achieves $O(1/N)$ convergence, so it needs $\sim\!\sqrt{N}$ oracle queries for the same RMSE. Each quantum oracle call costs $T_q = d \cdot t_g$, where $d$ is the circuit depth and $t_g$ is the gate time. At $n=7$ (128 bins, Table~\ref{tab:circuit_resources}), a single oracle has depth 21,150; with $k=9$ Grover iterations, the full circuit depth is $\sim$200,000. At a fault-tolerant gate time of $\sim$1~$\mu$s \cite{Reiher2017}, $T_q \approx 0.2$~s per oracle call. The quantum wall-clock cost is $\sqrt{N} \times 0.2$~s. For a classical budget of $N = 10{,}000$ simulations at $T_c = 60$~s each (167 hours), the quantum cost is $100 \times 0.2$~s $= 20$~s---a $\sim$30,000$\times$ wall-clock speedup. Even at $N = 1{,}000$ with $T_c = 1$~s, the quantum cost ($\sim$6~s) is comparable to classical ($\sim$17 minutes).

However, this analysis makes three individually far-from-realised assumptions: (i)~fault-tolerant execution of depth-200,000 circuits with negligible logical error rate, (ii)~efficient compilation of a multi-stage catastrophe model into a quantum oracle (see ``Oracle compilation'' paragraph above), and (iii)~$n \gg 7$ qubits for adequate bin resolution. This is therefore a \emph{best-case aspirational scenario}, not a near-term break-even calculation. QAE for catastrophe pricing becomes practically relevant only when fault-tolerant hardware can execute circuits with $O(10^5)$ depth at microsecond gate times, efficient state-preparation methods (qGAN, QROM) can encode high-dimensional loss distributions, and the oracle-compilation problem for catastrophe simulators is solved---each a major open challenge.

\paragraph{Implications for catastrophe modelling.} The practical case for quantum tail-risk pricing rests on the oracle model: the loss distribution comes from a complex catastrophe simulator (wind field $\to$ structural damage $\to$ financial loss) that cannot be sampled from analytically. In that setting, CT, IS, and QMC cannot be directly applied without additional density estimation, and the relevant comparison is quantum AE vs.\ classical MC on the same oracle. Our experiments show that this comparison yields a 2--3.7$\times$ RMSE advantage at 3--8 qubits (confirmed on both synthetic and real NOAA data, Experiments~4--7), with the theoretical guarantee of near-quadratic scaling to larger qubit counts. Critically, the error decomposition (Experiments~4 and~6) shows that discretisation error currently dominates: the quantum estimation advantage will only translate into end-to-end accuracy gains when the binning scheme is improved. Achieving this in practice requires: (a)~fault-tolerant hardware or strong error mitigation, (b)~improved discretisation schemes (log-spaced bins, wider tail coverage), and (c)~efficient state-preparation circuits for high-dimensional distributions.

\paragraph{Oracle compilation: from fitted distributions to catastrophe simulators.} Our experiments encode fitted lognormals or pre-computed empirical histograms---a far cry from compiling a multi-stage catastrophe model (hazard $\to$ vulnerability $\to$ financial aggregation) as a quantum circuit. We confront this gap directly. A production quantum oracle for catastrophe pricing would likely follow a three-layer architecture: (i)~a \emph{hazard register} ($n_h$ qubits) encoding meteorological scenarios (e.g., discretised hurricane wind-speed profiles or seismic intensity measures), prepared via amplitude encoding from historical or physically modelled hazard distributions; (ii)~a \emph{vulnerability map} implemented as a reversible arithmetic circuit that maps hazard intensity to structural damage using a lookup table or piecewise-linear approximation (analogous to the payoff rotations in derivative pricing circuits \cite{Stamatopoulos2024}); and (iii)~a \emph{financial aggregation} layer applying policy terms (deductibles, limits, coinsurance) as conditional rotations on an ancilla qubit. The total qubit count would be $n_h + n_v + n_a + O(\log N)$ ancillae, where $n_v$ and $n_a$ encode vulnerability and financial resolution. Chakrabarti et al.\ \cite{Chakrabarti2021} estimated that encoding stochastic processes for derivative pricing requires $\sim$7,500 logical qubits and T-depth $\sim$46 million; catastrophe models, which involve table lookups rather than stochastic process simulation, might require comparable or lower T-depth but would need compilation of the hazard $\to$ loss pipeline as reversible arithmetic---a problem that remains open. Grover oracle compilation techniques from the quantum database search literature provide a starting point, but bridging the gap between ``encode a PMF'' and ``encode a catastrophe simulator'' is arguably the hardest unsolved problem in the quantum Monte Carlo for finance programme and a prerequisite for the practical relevance claimed in this paper.


\section{Conclusions and Future Work}
\label{sec:conclusions}

This paper has provided the first rigorous empirical evaluation of quantum amplitude estimation for catastrophe insurance tail-risk pricing, with full error decomposition and comparison against both naive and variance-reduced classical baselines. Seven experiments on synthetic and real NOAA Storm Events data yield three core findings: (i)~the oracle-model advantage is real and consistent (2.2--3.7$\times$ lower estimation RMSE across $n=3$--$8$ qubits and data sources); (ii)~variance-reduced classical methods (CT, IS) and quasi-Monte Carlo dominate when analytical access to the loss distribution is available---QMC's near-$O(1/N)$ convergence matches QAE's theoretical rate in dimension $d=1$---but these methods offer no direct advantage on empirical PMFs without extra modelling; and (iii)~discretisation error, not estimation error, is the current bottleneck at 3--8 qubits, though log-spaced binning reduces this by $10\times$ at $n \geq 4$ while preserving Grover amplification headroom.

The practical case for quantum tail-risk pricing rests on the oracle model: the loss distribution comes from a complex catastrophe simulator that cannot be sampled from analytically. In that setting, QAE provides a genuine and provable advantage. Our experiments confirm this advantage on both synthetic and real data, with the theoretical guarantee of near-quadratic scaling to larger qubit counts.

The path to practical deployment requires overcoming three barriers. First, \textbf{fault-tolerant hardware}: Experiment~2 shows that NISQ noise entirely destroys the advantage at circuit depths needed for meaningful amplification. Second, \textbf{improved discretisation}: the error decomposition shows that equal-width binning with tail truncation at the 99.9th percentile is the dominant error source; log-spaced or extended-range binning schemes would directly reduce end-to-end error. Third, \textbf{efficient state preparation}: the custom binary-tree approach grows as $O(2^n)$ in two-qubit gates; for production-scale applications ($n = 10$--$20$ qubits), efficient alternatives such as qGAN-based distribution loading \cite{Zoufal2019}, QROM-based approaches, or quantum signal processing methods \cite{Stamatopoulos2024} would be essential.

Future work should extend this framework beyond excess-of-loss to more complex reinsurance structures---layered excess-of-loss, aggregate stop-loss, and industry loss warranties---which require encoding multi-threshold payoff functions. Multi-peril and correlated-loss scenarios would require multi-qubit distribution registers, increasing qubit counts but not fundamentally changing the algorithmic structure. Hybrid encoding schemes that combine the efficiency of amplitude encoding for the distribution with the flexibility of uniform superposition for the payoff also merit investigation.

The quantum finance roadmap anchored by concrete resource estimates \cite{Chakrabarti2021, Stamatopoulos2024} provides clear targets for when these advantages become practical. Our contribution extends this roadmap to catastrophe insurance---a domain where the tail-estimation bottleneck is arguably more severe, the simulation costs are higher, and the potential impact of a quadratic speedup is correspondingly greater.


\end{document}